\documentclass[10pt]{article}
\usepackage{bbm, amsmath, graphicx, mathrsfs, float, mathtools, cases, hyperref, color, cite, lineno, microtype, amssymb}
\DisableLigatures[f]{encoding = *, family = * }
\usepackage[labelfont=bf,labelsep=period,justification=raggedright]{caption}
\makeatletter
\renewcommand{\@biblabel}[1]{\quad#1.}
\makeatother

\date{}

\DeclareMathOperator*{\argmax}{arg\,max}

\newcommand{\1}{\mathbbm 1}

\newcommand{\yS}{\textbf{\small S}}
\newcommand{\yU}{\textbf{\small U}}
\newcommand{\yA}{\textbf{\small A}}

\newcommand{\dd}{\partial}
\newcommand{\sm}{\setminus}
\usepackage{pdfsync}

\begin{document}
\begin{flushleft}
{\Large
\textbf{Statics and dynamics of selfish interactions in distributed service systems}
}
\\
Fabrizio~Altarelli$^{1,2,\dag}$,
Alfredo~Braunstein$^{1,3,2\ddag}$,
Luca~Dall'Asta$^{1,2,\star}$\\
\small{
{\bf 1} Department of Applied Science and Technology and Center for Computational Sciences, Politecnico di Torino, Torino, Italy
{\bf 2} Collegio Carlo Alberto, Moncalieri, Italy
{\bf 3} Human Genetics Foundation, Torino, Italy
$\dag$ Presently at Capital Fund Management, Paris, France
$\ddag$ E-mail: alfredo.braunstein@polito.it
$\star$ E-mail: luca.dallasta@polito.it}
\end{flushleft}


\section*{Abstract}
We study a class of games which model the competition among agents to access some service provided by distributed \emph{service units} and which exhibit congestion and frustration phenomena when service units have limited capacity. We propose a technique, based on the cavity method of statistical physics, to characterize the full spectrum of Nash equilibria of the game. The analysis reveals a  large 
variety of equilibria, with very different statistical properties. Natural selfish dynamics, such as best-response, usually tend to large-utility equilibria, even though those of smaller utility are exponentially more numerous. Interestingly, the latter actually can be reached by selecting the initial conditions of the best-response dynamics close to the saturation limit of the service unit capacities. 
We also study a more realistic stochastic variant of the game by means of a simple and effective approximation of the average over the random parameters, showing that the properties of the average-case Nash equilibria are qualitatively similar to the deterministic ones.

\section*{Introduction}

Health care, education and communications are public sectors in which administrations face the problem of organizing distributed service systems. Service provision is {\em distributed} because a (possibly large) number of post-offices, hospitals, or libraries are maintained over the territory in order to allow all citizens to access the service.  Moving from public economics to information technology and computer networks, one can easily find other examples of distributed service provision: file-storage and file-sharing systems maintain servers (or mirrors) arranged all around the world to offer faster download to clients, while large wireless networks in airports and university campuses count dozens of access points ensuring connection to thousands of potential users.

Public service can be provided by a unique central administrator, such as the government or a monopolist, as well as by multiple competing providers. Although it is known that public goods such as education \cite{levy_politics_2005}, infrastructures \cite{lijesen_public_2007} and healthcare \cite{culyer_healthcare_1990} can be supplied by a market system, government policy interventions may be necessary in order to guarantee equity and quality of service, and to promote competitive conditions. 
Instabilities and inefficient outcomes are natural consequences of the awkward {\em non-exclusive} ownership nature of public goods,  a feature that has made public goods provision an active field of study in economics for decades. Users tend in fact to take advantage of public goods without contributing sufficiently to their creation and supply.  Without a mechanism of contribution, either voluntary or based on taxation, public service provision cannot be sustained, leading to the collapse of the system ({\em tragedy of the commons}) \cite{ostrom_tragedy_2008}. Although central in public service provision, this free-riding behavior is not the only source of inefficiency. Another relevant problem afflicting systems of non-excludable goods is {\em congestion} due to limited resources, a situation in which the service is not provided to all users or it does not give them equal satisfaction. When the service provision is free, or indirectly funded by some general taxation system, the system is not affected by free-riding phenomena, but inefficiency due to congestion effects can still be present. 

In this paper we will consider this simplified scenario, in which {\em users} have free access, through one of a set of distributed {\em service units}, to a limited amount of resources that is managed by a unique central authority, i.e. the {\em service provider}. 
In this respect the problem looks like a standard optimization problem, with the administrator seeking the optimal resource allocation, in which all users are served and the load on the units is balanced minimizing the costs.
On the other hand, the strategic, non-cooperative character of the problem is evident from the fact that users are self-interested agents. Every user wants to be served from the service unit that is most convenient to her, because of a smaller load, geographical proximity or a higher quality of service. Since resources are limited, the individual utilities of the users depend, directly or indirectly on the usage that others do of the same service unit.
Service units have typically different quality of service, and the latter may also depend on the workload of the unit at the time of service.  For instance, when waiting lists become too large, choosing the  best hospital or the most efficient public office could become inconvenient. In file-sharing systems the time required to download files depends on the number of requests to the same server. When too many users download from the same server, the quality of service deteriorates. Wireless access points serve users according to a round robin, providing one opportunity to transmit at each user during a cyclic time frame of finite  duration. This physical constraint imposes a limitation on the number of users that can be connected to a single access point at the same time. Similar phenomena of ``congestion'' occur in every distributed service provision system and make the organizational problem a game theoretic one, in which the non-cooperative behavior of the users causes degradation with respect to a centralized optimal solution.

The outcome of any strategic interaction among rational agents can be predicted and classified using the concept of {\em Nash equilibrium}, which describes a situation in which no player has incentive to unilaterally deviate from the chosen strategy profile. As a result of the simultaneous maximization of all individual utilities, a game can admit more than one Nash equilibrium. The existence of multiple Nash equilibria is particularly common in strategic decision problems in disordered and networked systems, where the number of different Nash equilibria can even scale exponentially with the number of players. As many other systems in computer science \cite{nisan_algorithmic_2007} and network economics \cite{jackson_social_2010}, distributed service provision systems can admit a large multiplicity of Nash equilibria.

Whenever multiple equilibria exist, all of them are equally rational and there is no a priori way to state which one would be chosen by the agents. This lack of predictive power is usually resolved by introducing some refinement of the concept of Nash equilibrium or some criterion of equilibrium selection. A reasonable assumption is 
that the selected Nash equilibrium should be the outcome of a dynamical process in which agents may interact several times. However, no dynamical rule is universal and different ones lead to equilibria with possibly very different properties. 
A detailed knowledge of the equilibrium landscape is crucial to devise reasonable criteria of equilibrium selection and to discriminate between the outcomes of different dynamical rules. 
Such information can be then exploited to design effective self-enforcing mechanisms, for example by means of incentives, to move the system away from bad equilibria towards the most efficient ones. 

It was recently proved possible to investigate the whole landscape of Nash equilibria in multi-agent games, such as public goods games on networks \cite{dallasta_optimal_2011,dallasta_statistical_2009,ramezanpour_statistical_2011}, by mapping the equilibrium condition on a constraint-satisfaction problem and then analyzing it using efficient message-passing algorithms based on the cavity method from statistical physics \cite{mezard_bethe_2001, mezard_cavity_2003, mezard_information_2009}.
Here we adopt this approach to study the equilibrium properties in a simple model of distributed service provision. We use such information to understand how efficient (i.e. with large aggregate utility) Nash equilibria obtained from best-response are compared to those obtained using different dynamical rules and how much the answer depends on the initial conditions.
Moreover, the real self-organization processes of the agents are not exactly best-response processes and their details are normally not known, therefore a complete analysis cannot be restricted to a single dynamics. We bring evidence of the richness of the equilibrium landscape by describing the full set of Nash equilibria using a statistical mechanics analysis and comparing it with the typical fixed-points of different dynamics. We also study the effects of correlating the users' utilities with the loads they bring to the system. Finally we generalize the analysis to a stochastic case, in which the agents are present with a given probability. To do this, we introduce a new algorithmic approximation technique to perform the required average over the realizations of the stochastic parameters on single instances.
 
\section*{Related Work}

Congestion effects have been isolated and studied by means of game-theoretic models introduced by Rosenthal \cite{rosenthal_class_1973} and called {\em congestion games}. These are games in which players use resources from a common pool, and the payoff of each player depends on the total {\em load} present on the resources she chooses. The load on each resource is thus the relevant quantity defining congestion effects and it is function only of the number of agents using that resource. Rosenthal showed that pure Nash equilibria always exist for such games \cite{rosenthal_class_1973}. {\em Weighted} congestion games, with player-specific utility functions, were introduced by Milchtaich \cite{milchtaich_congestion_1996}, and recently studied in several contexts \cite{ackermann_pure_2006,panagopoulou_algorithms_2007,mavronicolas_congestion_2007,bhawalkar_weighted_2010,gourves_congestion_2012}. In particular, a class of congestion games with capacitated resources, where each resource is associated with a capacity level, representing the maximum number of users that such a resource may simultaneously accommodate, was recently investigated in \cite{gourves_congestion_2012}.

There are innumerable examples of congestion games in relevant socio-economic and technological systems, including network routing \cite{koutsoupias_worst-case_1999,roughgarden_how_2002,roughgarden_bounding_2004,czumaj_tight_2007}, bandwidth and spectrum sharing in communication networks \cite{altman_survey_2006,liu_spectrum_2008} and market-entry problems \cite{rhim_competitive_2003,anderson_participation_2007}. Minority games \cite{challet_minority_2013} can also be interpreted as a special class of congestion games.  In some of these applications, the game is {\em non-atomic} or continuous, namely it is defined on an infinitely large population of agents, and the load function depends of the density of agents that use that resource in the population. 

Since the typical setup of a congestion game corresponds to the selfish, autonomous, generalization of a traditional resource allocation problem, it is thus important to quantify how much the efficiency of a system degrades due to the selfish behavior of its agents, i.e. the difference between optimal centralized solution of the allocation problem and the selfish ones. To this purpose, 
computer scientists introduced the concepts of {\em Price of Anarchy} \cite{koutsoupias_worst-case_1999} and {\em Price of Stability} \cite{anshelevich_price_2008}.  The Price of Anarchy is the ratio between the utility of the social optimum and the utility of the worst equilibrium, while the Price of Stability is the ratio between the utility of the social optimum and that of the best equilibrium. In fact, these quantities only give bounds on the equilibrium landscape, providing no further information on its structure. 
In the continuous game setup, Roughgarden and Tardos proved that, for linear load functions, the price of anarchy of  systems is bounded and not large, demonstrating that selfish behavior does not necessarily lead to very inefficient outcomes  \cite{roughgarden_how_2002,roughgarden_bounding_2004}. Similar bounds were obtained for atomic congestion games by Suri et al. \cite{suri_selfish_2004}. For non-atomic congestion games, Correa et al. \cite{correa_selfish_2004} showed that in capacitated systems, the price of anarchy is not bounded anymore, although the best Nash equilibrium can be as efficient as for the model without capacity constraints.

Rosenthal's proof that pure strategy Nash equilibria always exist in congestion games was based on the construction of a potential function. {\em Potential games} \cite{monderer_potential_1996} are games that admit a potential function with the property that an improvement of an individual player decreases the potential by exactly the same amount as the player's cost. Nash equilibria are thus the local minima of such a potential and it is possible to show that the optimal assignment is also a Nash equilibrium. As the set of strategies is finite and the potential function increases during the best-response dynamics, the latter always converges to a Nash equilibrium. Monderer and Shapley \cite{monderer_potential_1996} showed that any potential game can be represented in form of a congestion game. 

\section*{The model}

\subsection{The service provision game} \label{sec:the_service_provision_game}

\begin{figure}[t]
\begin{center}
	\includegraphics[width=0.6\columnwidth]{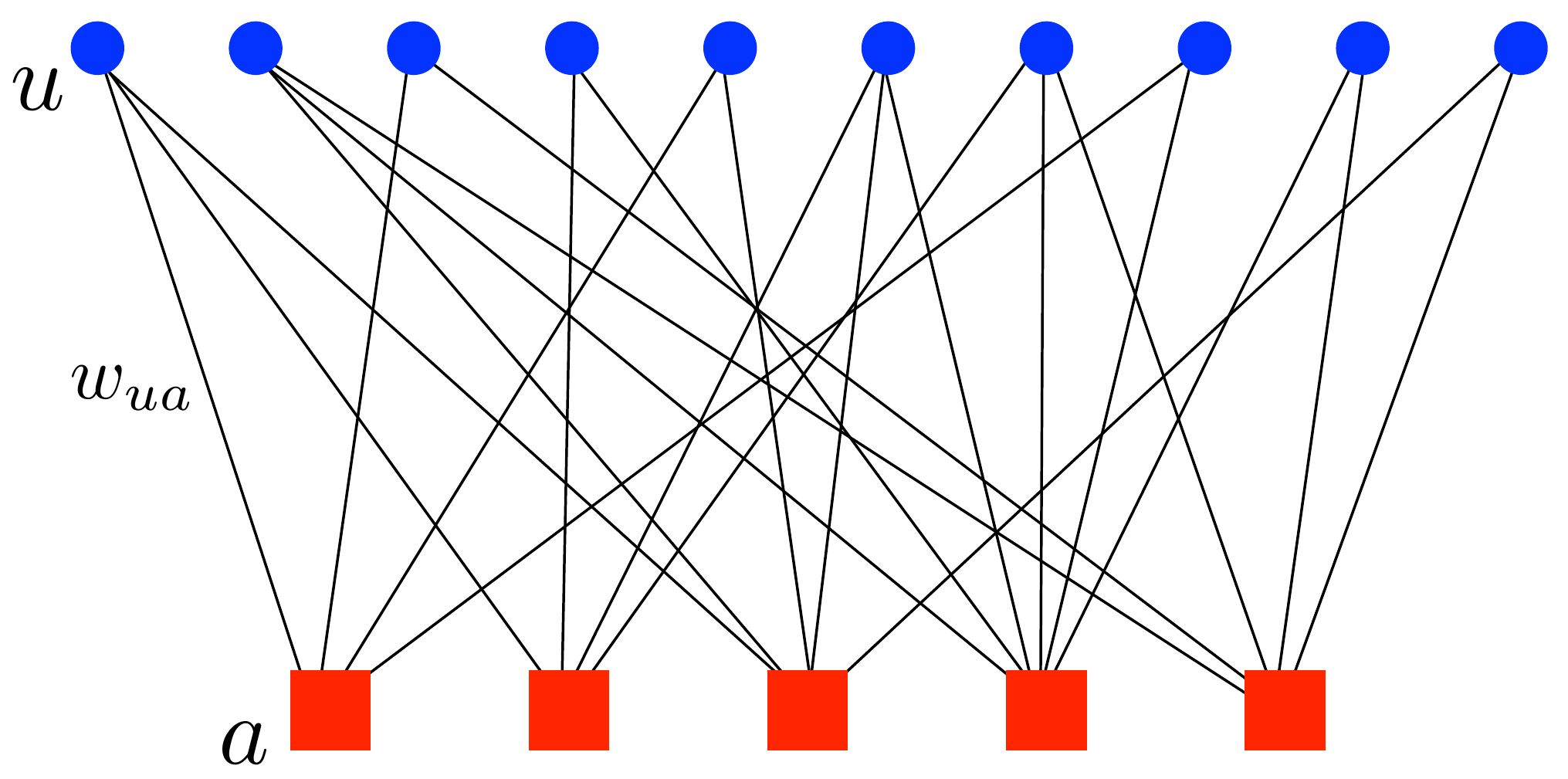}
\caption{\label{fig:bg} Bipartite graph representing an instance of the service provision game, with weights $w_{ua}$ between users $u \in \mathsf{U}$ and service units $a \in \mathsf{S}$.}
\end{center}
\end{figure}

In the service provision game, there are two types of entities: users and service units. Each user benefits from being serviced by one service unit, with each service unit providing her a different utility. A user would prefer being serviced by the unit that provides her the largest utility. However, service units have finite capacity, so in certain cases the first choice for a given user can be unavailable.
An instance of the service provision game is represented by a bipartite graph $\mathsf{G}=(\mathsf{U},\mathsf{S};\mathsf{E})$ as in Fig.~\ref{fig:bg}, where $\mathsf{U} = \{1, \dots, N\}$ is the set of users, $\mathsf{S} = \{1, \dots, M\}$ is the set of service units, and an edge $(ua)$ is present in $\mathsf{E}$ if and only if the service unit $a$ is accessible (but not necessarily available) to the user $u$. We associate to each edge $(ua) \in \mathsf{E}$ a positive weight $w_{ua}$, which represents the {\em load} placed on service unit $a$ by user $u$, and a positive quantity $v_{ua}$ representing the (utility) {\em value} that $u$ gives to the service provided by $a$. We also associate a capacity $C_a$ to each service unit $a$, representing the maximum load it can serve. In a general setup, values and weights are heterogeneous, as the same user $u$ may provide a different load $w_{ua}$ to different service units $a$ and obtain different levels of satisfaction $v_{ua}$.
The two quantities on the same edge $(ua)$ can be correlated (either positively or negatively) or uncorrelated; all cases are of interest. 

The {\em action} of a user $u$ corresponds to the choice among the $M_u$ service units accessible to user $u$. For each edge $(ua)$, we introduce a binary variable  $x_{ua} \in \{0,1\}$, such that $x_{ua} = 1$ iff user $u$ is served by service unit $a$. The action of user $u$ is given by the binary vector ${\bf x}_u= (x_{u1}, x_{u2}, \dots, x_{uM_u})$ in which at most one component can be equal to $1$, i.e. it satisfies the condition 
\begin{subequations} \label{eq:constraints1}
\begin{align}
  \sum_{a \in \dd u} x_{ua} &\leq 1 & (\forall u \in \mathsf{U}) \,,\label{eq:matching}
\end{align}
where $\dd u \subset \mathsf{S}$ denotes the neighbors of $u$ on the graph. The capacity constraints can be written as
\begin{align} \label{eq:capacity}
  \sum_{u \in \dd a} x_{ua} w_{ua} &\leq C_a &(\forall a \in \mathsf{S}).
\end{align}
\end{subequations}
The utility for user $u$ is simply
\begin{align}
U_u = 
  \begin{cases}
        \sum_{a \in \dd u} x_{ua} v_{ua} & \text{if \eqref{eq:constraints1} are satisfied,} \\
      -\infty & \text{otherwise}.
   \end{cases}
\end{align}

An action profile ${\bf x} = \{{\bf x}_u\}_{u  \in \mathsf{U}}$ is a (pure) {\em Nash equilibrium} of the service provision game if no user can increase her utility by unilaterally switching to a different service unit, i.e. for each user $u$
\begin{equation}
U_u({\bf x}) \geq \max_{{\bf x'}_u} U_u({\bf x'}_u, {\bf x}_{\sm u})
\end{equation}
where $U_u({\bf x'}_u, {\bf x}_{\sm u}) = U_u({\bf x}_1, \dots, {\bf x}_{u-1}, {\bf x'}_u,{\bf x}_{u+1} \dots, {\bf x}_{N})$. 

Let us call $X^N$ the action space. A game is an {\em exact potential game} if there exists a potential function $V: X^N \to \mathbbm{R}$, such that for each user $u$
 \begin{equation}
 U_u({\bf x}_u, {\bf x}_{\sm u}) - U_u({\bf x'}_u, {\bf x}_{\sm u}) = V({\bf x}_u,{\bf x}_{\sm u}) - V({\bf x'}_u,{\bf x}_{\sm u}).
 \end{equation}
 The service provision game is a potential game that admits the aggregate utility $U = \sum_u U_u$ as an exact potential $V$.  This statement is easy to prove because the benefit that a user receives for being connected to a service unit does not directly depend on the actions of the others but only on the availability of the unit itself. It follows that when a user moves to a different service unit, her action does not affect the utilities of all the other users, even of those that are disconnected from the system. In this respect, a better measure of social welfare should take into account both the aggregate utility $U$ and the number $D$ of disconnected users, e.g. in a linear combination $U - \alpha D$. The two definitions of utility $U$ and $U'=U-\alpha D$ can be reconciled by considering a slightly modified game, in which we add a virtual service unit for each user with unlimited capacity and negative utility $-\alpha$, in a game in which each user is forced to be connected to exactly one unit (i.e. there is an equal sign in equation \ref{eq:matching}). Because of the negative contribution to the utility, this new ``{\em unservice unit}'' will be chosen only as a last resort, when all real units are saturated and the user would be otherwise disconnected. In this modified game, users are always connected (but connection to the {\em unservice unit} represents disconnection in the original game) and the total utility is $U' = U - \alpha D$. We do not use this interpretation in the equations, but it can be useful to keep in mind.
In any case, if a user can improve her utility with a feasible move, this will not affect the social welfare enjoyed by other users, and the total social welfare will necessarily improve, which means that a configuration which is not a Nash equilibrium cannot be the social optimum. The Nash equilibria of the service provision game are in one-to-one relation with the local maxima of the potential $V$, i.e. of the aggregate utility $U$. This can be verified by defining the {\em best response} relations. 
For a user $u$, the  best response to the actions of the other users consists in choosing the action that maximizes her own utility given the choice of the others, that is ${\bf x}_u$ is the best response for $u$ to the remaining action profile ${\bf x}_{\sm u}$ if  
\begin{equation}
{\bf x}_{u} = \argmax_{{\bf x'}_u} U_u({\bf x'}_u, {\bf x}_{\sm u}).
\end{equation}
In a potential game with discrete actions and finite strategy space, the potential does not decrease during the iteration of best-response reactions, and the latter always converge to a pure Nash equilibrium in a finite number of steps \cite{monderer_potential_1996}. Because of this property, the path in the action space generated by the iteration of the best response relations is often called the {\em improvement path}.

\subsection{Example} \label{sec:example}

\begin{figure}[t]
\begin{center}
	\includegraphics[width=0.4\columnwidth]{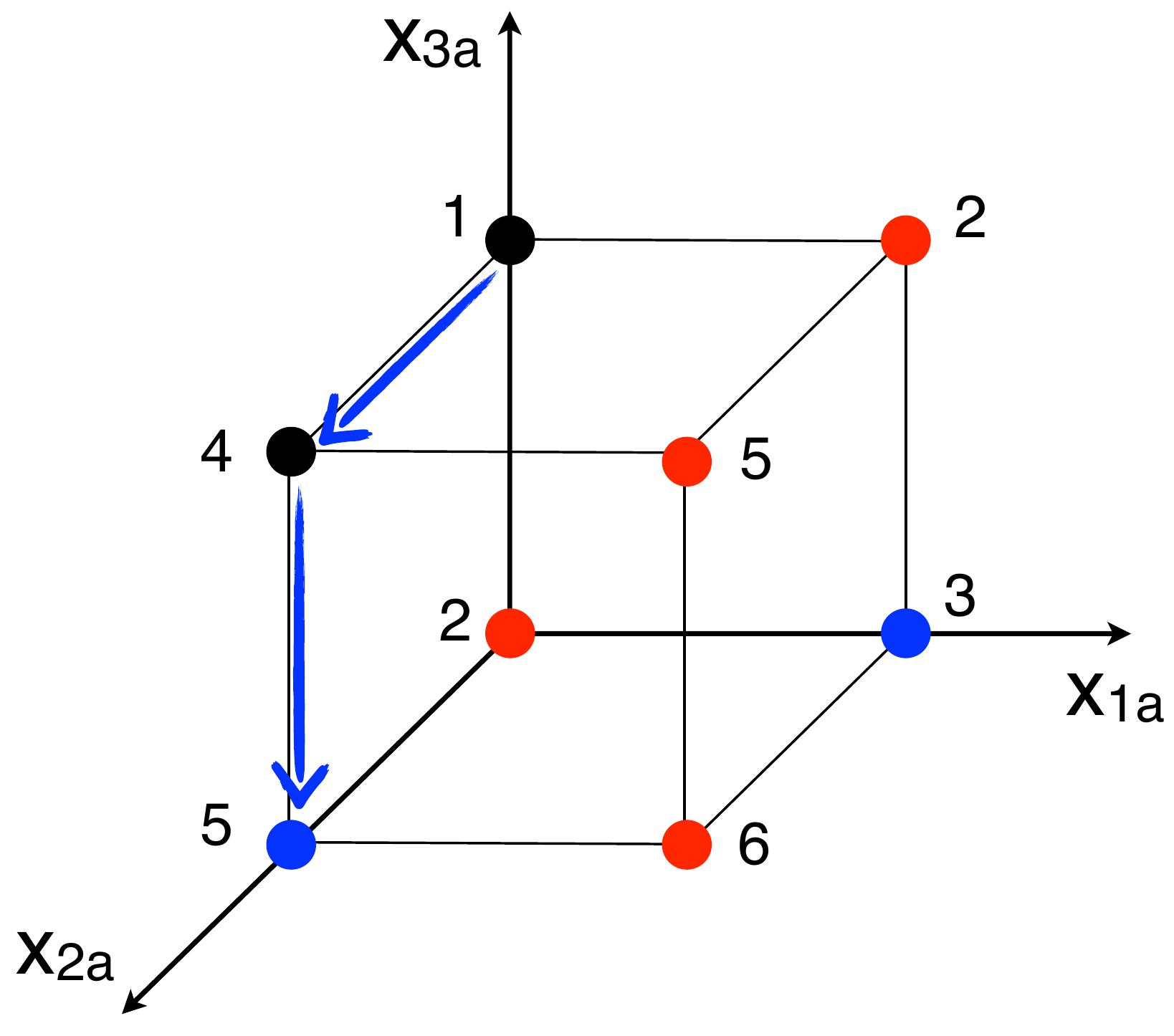}
\caption{\label{fig:example} Binary representation $(x_{1a},x_{2a}, x_{3a})$ of the action space for the example presented in Section~\ref{sec:the_service_provision_game} with 3 users and 2 service units. Each vertex corresponds to a possible configuration: non-feasible configurations, that violate capacity constraints, are marked in red, Nash equilibria are marked in blue. The value of the aggregate utility is also reported for each configuration. The blue arrows indicate an improvement path obtained by best response from $(0,0,1)$ to the Nash equilibrium $(0,1,0)$.}
\end{center}
\end{figure}

Let us review all these properties in a simple example of service provision game composed of three users $\mathsf{U} = \{1,2,3\}$ and two service units $\mathsf{S} = \{a,b\}$. The weights are $w_{1a} = 3, w_{1b} = 1$,$w_{2a} = 1,w_{2b} = 2$, and $w_{3a} = 1,w_{3b} = 2$, while the values given to the service are  $v_{1a} = 2, v_{1b} = 1$,$v_{2a} = 3,v_{2b} = 0$, and $v_{3a} = 0,v_{3b} = 1$. Service units have capacities $C_a = 3, C_b = 4$. In this example, the values of weights and capacities are defined in such a way that users can always be connected to one of the two service units. In this case, for each user $u$ either $x_{ua}$ or $x_{ub}$ will be equal to 1 and the action profile simplifies considerably. Since $x_{ub} = 1- x_{ua}$, the aggregate utility can be written as 
\begin{equation}
U = \sum_{u=1}^{3} \sum_{a}\left( v_{ua} x_{ua} + v_{ua} (1-x_{ua}) \right) = 2 + x_{1a} + 3x_{2a} - x_{3a},
\end{equation}
and the capacity constraints on the loads of the service units are
\begin{subequations}
\begin{align}
\ell_a & = \sum_u w_{ua} x_{ua} = 3 x_{1a} + x_{2a} + x_{3a}  \leq 3 \\
\ell_b & = \sum_u w_{ub} x_{ub} = 5 - x_{1a} - x_{2a} - x_{3a}  \leq 4. 
\end{align}
\end{subequations}
In the reduced action space $(x_{1a}, x_{2a}, x_{3a})$, the feasible action profiles are $(1,0,0)$, $(0,1,0)$, $(0,0,1)$, $(0,1,1)$, and the Nash equilibria are $(0,1,0)$ and $(1,0,0)$ (see Fig.\ref{fig:example}). Suppose the system is in the feasible configuration $(0,0,1)$, with aggregate utility $U=1$ and let the users perform best response. User $1$ will not move from unit $b$ to unit $a$, although the latter could provide a better service, because of the capacity constraint on $a$; user $3$ will not move from $a$ to $b$ for the same reason. For user $2$, it is instead convenient to leave service unit $b$ and connect to $a$,  because $v_{2a} = 3 > v_{2b} = 0$ and the weight $w_{2b} = 1$ will not cause any capacity violation on $a$. The new configuration $(0,1,1)$ has aggregate utility $U=4$ but it is not a Nash equilibrium. Indeed, user $3$ can now increase her own utility by switching from unit $a$ to unit $b$, which has now sufficient spare capacity. The action profile $(0,1,0)$ is a fixed-point for the best response dynamics and a pure Nash equilibrium of the game. Notice that the Nash equilibrium is also a local maximum for the aggregate utility ($U=5$). There is another Nash equilibrium in which user $1$ is connected to unit $a$, while both 2 and 3 use unit $b$. This Nash equilibrium has a low aggregate utility $U=2$ but it saturates the capacities of both units (loads $\ell_a = 3$ and $\ell_b = 4$). Interestingly, the best Nash equilibrium induces loads $\ell_a = 1$ on unit $a$ and $\ell_b = 3$ on unit $b$, that are instead far below the capacity limits. We will see that this phenomenology is a general feature of equilibria in the service provision game.

When there are many users and service units, with different weights, service values and capacities, the number of Nash equilibria grows exponentially with the size of the instance and they will show a wide spectrum of properties. In Section~\ref{sec:mapping_csp} and Section~\ref{sec:bp_equations}, we will put forward a method, based on statistical mechanics techniques, to study the properties of all Nash equilibria, classifying them depending on different quantities. In many realistic situations, however, the instance of the game theoretic problem changes over time, because the agents could follow very complex temporal activity patterns. An example is given by wireless service provision: a provider can have information of all users potentially connected to the network, but it will not be able to anticipate the exact time at which they will be connected and the duration of the connections.
One can imagine that agents could leave the system and come back, using different service units depending on their preference and the current availability. In the absence of any precise information on the dynamics of the agents, one could be tempted to use a standard approach in games with {\em incomplete information}: the lack of information about the complexity of agents' activity is summarized into a set of stochastic parameters $\{ t_u\}_{u \in \mathsf{U}}$. If $t_u = 0$ the user $u$ is \emph{inactive} (meaning that $u$ doesn't participate to the game), whereas $t_u =1$ if she is \emph{active} (meaning that $u$ participates to the game). The probability that user $u$ is active ($t_u = 1$) is $p_u$, which we assume to be known, and it is independent on the state of other users. 
However, under the standard assumptions of incomplete information, agents do not know the exact realization of the stochastic parameters, but only their probability distribution. Agents maximize an expected utility and the notion of Nash equilibrium is replaced by that of Bayesian Nash equilibrium \cite{jackson_social_2010}. 
In the present setup, the major drawback of the formulation with incomplete information is that it does not describe correctly the behavior of the users. In realistic situations, users connect to the system and possibly rearrange their decisions on the base of the current instance of the problem, until they reach an equilibrium. In other words, we expect that users play a game of complete information on the deterministic instance corresponding to a single realization of the stochastic parameters $\{ t_u\}_{u \in \mathsf{U}}$. The knowledge of the probabilities $p_u$ is instead relevant in order to sample over many realizations of the stochastic parameters and evaluate the average properties of the Nash equilibria of the game. In Section \ref{sec:mirror}, we will show that our methods can be generalized in order to average the properties of the Nash equilibria over the realization of the stochastic parameters without resorting to sampling techniques. 

\setcounter{subsection}{0}
\section*{Methods}

\subsection{Representation as a constraint-satisfaction problem}\label{sec:mapping_csp}

\subsubsection*{The deterministic case}
The topology of the graph together with the values of the parameters $\{w_{ua},\, (ua) \in \mathsf{E}\}$, $\{v_{ua},\, (ua) \in \mathsf{E}\}$ and $\{C_a,\, a \in \mathsf{S}\}$ completely define an instance of the game. In the following, we shall assume that all the weights, utilities and capacities are positive integers. In the following we will show that the Nash equilibrium conditions can be mapped on the solutions of a constraint-satisfaction problem. In order to do that, we introduce a convenient set of variables $y_{ua} \in \{ \yU, \yA, \yS\} $ associated to the edges $(ua) \in \mathsf{E}$ of the graph, representing both the choice of user $u$ and the availability of service unit $a$ as follows:
\begin{align}
  y_{ua} =
    \begin{cases}
      \yU & \text{if $a$ is unavailable to $u$,} \\
      \yA & \text{if $a$ is available to $u$, but $u$ is not served by $a$,} \\
      \yS & \text{if $u$ is served by $a$.}
    \end{cases}
\end{align}
In order for a configuration of the constraint satisfaction problem to be a valid configuration in the service provision game, it must satisfy the following constraints. First, each user can be served by at most one service unit
\begin{subequations} \label{eq:constraints}
\begin{align} \label{eq:constraint_u_sum}
  \sum_{a \in \dd u} \1[y_{ua} = \yS] &\leq 1 & (\forall u \in \mathsf{U})
\end{align}
where $\1[\textsl{proposition}]$ is the indicator function for \textsl{proposition}, which is equal to 1 if \textsl{proposition} is true and 0 otherwise. Second, the total load on each service unit cannot exceed its capacity:
\begin{align} \label{eq:constraint_s_sum}
  \sum_{u \in \dd a} w_{ua} \1[y_{ua} = \yS] &\leq C_a & (\forall a \in \mathsf{S}) \,.
\end{align}
Third, a service unit $a$ is available to a user $u$ (not currently served by $a$) if and only if $a$ has a spare capacity sufficient to serve $u$:
\begin{align} \label{eq:constraint_s_available}
  \Big\{y_{ua} = \yA \Big\} \Leftrightarrow \Big\{ w_{ua} + \textstyle \sum_{v \in \dd a \sm u} w_{va} \1[y_{av} = \yS] \leq C_a \Big\} \wedge \Big\{y_{ua} \neq \yS\Big\}  && (\forall (ua) \in \mathsf{E}) \,.
\end{align}
A valid configuration is a Nash equilibrium if it satisfies the further condition that each user is served by the best available service unit, or equivalently that if a service unit $a$ is available to user $u$ but not used by $u$, then $u$ must be served by some other service unit $b$ with a utility at least as large as $a$'s:
\begin{align} \label{eq:constraint_u_optimal}
  \Big\{y_{ua} = \yA \Big\} \Leftrightarrow \Big\{ \exists b \in \dd u:\, \{y_{ub} = \yS \} \wedge \{v_{ub} \geq v_{ua}\} \Big\} \wedge \Big\{ y_{ua} \neq \yU \Big\} && (\forall (ub) \in \mathsf{E}) \,.
\end{align}
\end{subequations}
The sets of conditions \eqref{eq:constraints} are the hard constraints characterizing the solutions of the constraint-satisfaction problem under study, that are the Nash equilibria of the service provision game. 

\subsubsection*{The stochastic case}
An instance of the stochastic service provision game is completely determined by the topology of the graph and the values of the parameters $\{w_{ua},\, (ua) \in \mathsf{E}\}$, $\{v_{ua},\, (ua) \in \mathsf{E}\}$, $\{C_a,\, a \in \mathsf{S}\}$ and $\{p_u,\, u \in \mathsf{U}\}$. 
A configuration of the stochastic game is described by the pair $(t, y)$ where $t = \{t_u,\, u \in \mathsf{U}\}$ represents the activity of users, and $y = \{y_{ua},\, (ua) \in \mathsf{E}\}$ represents their actions. The conditions \eqref{eq:constraints} for $y$ to be a Nash equilibrium remain unchanged, except for the first one, which becomes
\begin{align}
  \sum_{a \in \dd u} \1[y_{ua} = \yS] \leq t_u && (\forall u \in \mathsf{U}) \,.
\end{align}

\subsection{Probability measure over the Nash equilibria}
In this constraint-satisfaction representation of the game-theoretic model, the set of Nash equilibria is in one-to-one correspondence with the configurations of discrete variables that satisfy a set of local constraints. Although solving contraint satisfaction problems could in general be computationally difficult, we have seen in the model description that this problem corresponds to a Potential Game, and finding a solution is computationally easy. We will show in the subsection \ref{SubSec:dynamics} of Results some explicit examples of algorithms (e.g. the Greedy dynamics) that can find a solution using a number of operations that scales polynomially with the number of variables. 

Still, at least two computational difficulties remain for the study of this game. The first is that some of the equilibria could be hard to find, and it could be the case that only a subset of ``easy'' equilibria are reachable by simple algorithms. Second, for our analysis, we aim at characterizing aggregated properties of the space of equilibria. Even in the unlikely case in which all equilibria are ``easy'', enumerating all of them could be computationally unfeasible when the number of variables is large. In the following we describe statistical physics methods to investigate their statistical properties (without resorting to an explicit enumeration), following a general approach to the study of constraint-satisfaction problems based on the cavity method \cite{mezard_bethe_2001, mezard_cavity_2003, mezard_information_2009}. Interestingly, this method also provides very efficient algorithms to find Nash equilibria with typical and non-typical properties.

Although all Nash equilibria are a priori equally rational,  we expect that they could have different properties, therefore we shall be interested to compute the average value $\overline O$ of some extensive observable $O(y)$ with a uniform measure over the Nash equilibria:
\begin{align}
  \overline O &= \frac 1 {\mathcal N} \sum_y O(y) \mathscr C(y)
\end{align}
where $\mathcal N$ is the total number of Nash equilibria and $\mathscr C(y)$ is equal to 1 if the conditions \eqref{eq:constraints} are satisfied and 0 otherwise, so that $\mathcal N = \sum_y \mathscr C(y)$.
Some interesting observables will be the aggregate utility
\begin{align}
  U(y) = \sum_{(ua) \in \mathsf{E}} v_{ua} \1[y_{ua} = \yS] \,,
\end{align}
the total number of users who are disconnected (i.e. who are not served by any service unit)
\begin{align}
  D(y) = \sum_{u \in \mathsf{U}} \1 \left[ \sum_{a \in \dd u} \1[y_{ua} = \yS] = 0 \right] \,,
\end{align}
or the aggregate unutilized (spare) capacity
\begin{align}
  C^*(y) = \sum_{a \in \mathsf{S}} \left\{ C_a - \sum_{u \in \dd a} w_{ua} \1[y_{ua} = \yS] \right\} = \sum_{a \in \mathsf{S}} C_a - L(y)
\end{align}
where $L(y)$ is the aggregate load.

The uniform measure assigns the same finite weight to all configurations of variables corresponding to Nash equilibria and zero to all remaining ones (as they violates some of the best-response constraints). In order to characterize the Nash equilibria which correspond to a given value of some observable $O(y)$, we can replace the uniform measure with an exponential family, a Gibbs measure 
\begin{align} \label{eq:Gibbs_mu}
  P(y | \mu) = \frac 1 {Z(\mu)} \exp\{ \mu O(y) \} \mathscr C(y)
\end{align}
where $Z(\mu) = \sum_y \exp\{ \mu O(y) \} \mathscr C(y)$ and the parameter $\mu$ controls the weight to be assigned to Nash equilibria with different values of the 
the observable $O(y)$. 
Notice that for $\mu = 0$ we recover the uniform distribution over all Nash equilibria. Fixing the parameter $\mu$, the entropy
\begin{align}
  S(\mu) = - \sum_y P(y | \mu) \log P(y | \mu)
\end{align}
provides a measure of the number of Nash equilibria corresponding to the average value $\mathcal O(\mu)$ of $O(y)$, defined as
\begin{align}
  \mathcal O(\mu) = \sum_y O(y) P(y | \mu) \,.
\end{align}
Moreover, in the following, all quantities averaged over the distribution $P(y | \mu)$ will be denoted using a calligraphic fonts, such as $\mathcal{U}, \mathcal{D}, \mathcal{C}^*$.

When considering the average properties of an instance of stochastic service provision game, for example the average value $\left< O \right>$ of an observable $O(y)$, we must perform a double average: over all the Nash equilibria corresponding to a given realization of the parameters $t = \{t_u,\, u \in \mathsf{U}\}$, and over the realization of $t$:
\begin{align} \label{eq:avg_O_stochastic}
  \left< O \right> &= \sum_t P(t) \frac 1 {\mathcal N(t)} \sum_{y} O(y) \mathscr C(y, t)
\end{align}
where $\mathcal N(t)$ is the number of Nash equilibria corresponding to a given realization of $t$, and where $\mathscr C(y, t)$ is 1 if $y$ is a Nash equilibrium for $t$ and 0 otherwise.

\subsection{Belief Propagation equations} \label{sec:bp_equations}
\begin{figure}[t]
\begin{center}
	\includegraphics[width=0.8\columnwidth]{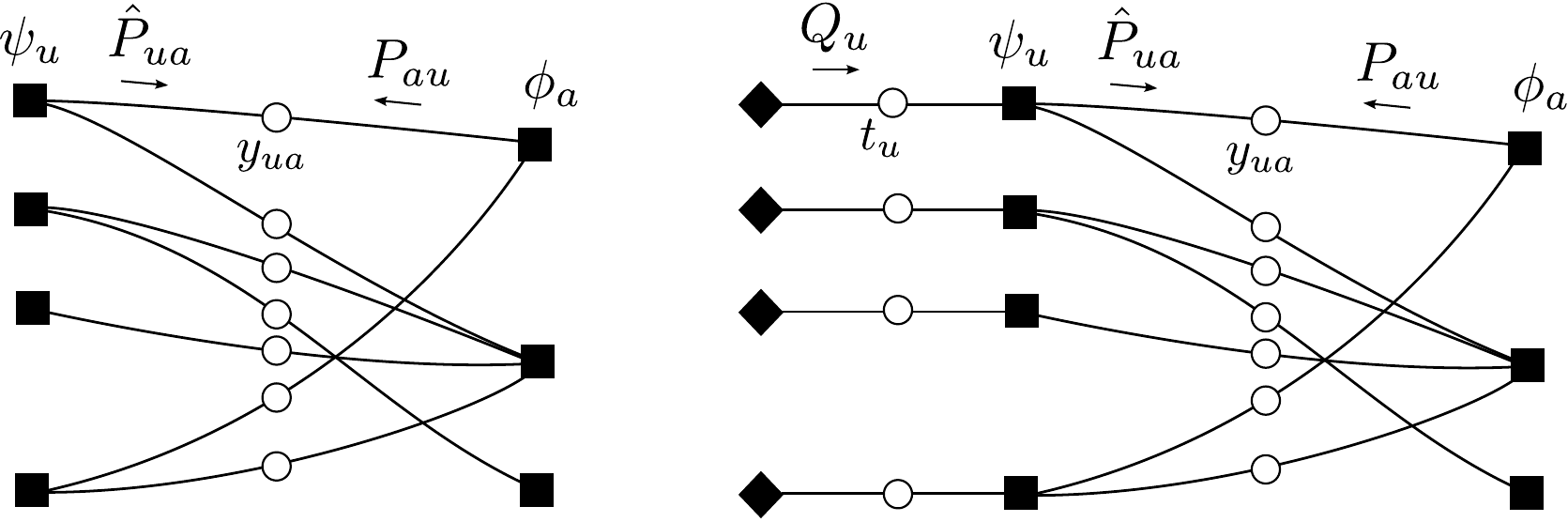}
\caption{\label{fig:factor_graphs} Factor graph representation. Left: For the deterministic case the set of constraints is $\mathscr C(y) = \prod_{u \in \mathsf{U}} \psi_u(y_u) \prod_{a \in \mathsf{S}} \phi_a(y_a)$. Right: For the stochastic case, the factor graph includes mirror nodes (black lozanges) and the corresponing messages $Q_u(t_u)$. The set of constraints is $\mathscr C(y,t) = \prod_{u \in \mathsf{U}} \psi_u(y_u, t_u)  \prod_{a \in \mathsf{S}} \phi_a(y_a)$.}
\end{center}
\end{figure}

The entropy, as well as the distribution over the Nash equilibria can be computed solving a set of local self-consistent equations for probability marginals, that are known as Belief Propagation (BP) equations. In this context, the probability measure \ref{eq:Gibbs_mu} is usually reinterpreted as a {\em graphical model} (see e.g. \cite{yedidia_understanding_2003}).
We denote $y_u = \{y_{ua},\, a \in \dd u\}$ the set of variables $y_{ua}$ on the edges incident on $u \in \mathsf{U}$, and similarly for $y_a = \{y_{ua},\, u \in \dd a\}$.
We consider a factor graph with the same topology as the bipartite graph $\mathsf{G} = (\mathsf{U}, \mathsf{S}; \mathsf{E})$ representing the instance, with factor nodes $\psi_u(y_u)$ associated to the users $u \in \mathsf{U}$ and enforcing the constraints \eqref{eq:constraint_u_sum} and \eqref{eq:constraint_u_optimal}, and factor nodes $\phi_a(y_a)$ associated to the service units $a \in S$ and enforcing the constraints \eqref{eq:constraint_s_sum} and \eqref{eq:constraint_s_available} (Figure \ref{fig:factor_graphs}, left-hand panel).

We introduce the BP messages $P_{au}(y_{ua})$ traveling on the edge $(ua) \in \mathsf{E}$ from the service unit $a$ to the user $u$, and $\widehat P_{ua}(y_{ua})$ traveling in the opposite direction. Notice that since all the variables have connectivity 2 there is no difference between variable-to-factor and factor-to-variable messages. In the following we derive the BP equations for the uniform distribution over all Nash equilibria, i.e. setting $\mu = 0$ in \eqref{eq:Gibbs_mu}. It is straightforward to introduce a bias related to the value of any extensive observable $O(x)$ as in \eqref{eq:Gibbs_mu}.
Formally, the BP equations can be written as
\begin{subequations}
\begin{align} \label{eq:BP_det_naif}
  P_{au}(y_{ua}) &= \frac 1 {z_a} \sum_{\{y_{va},\, v \in \dd a \sm u\}} \phi_a(y_a) \prod_{v \in \dd a \sm u} \widehat P_{va}(y_{va}) \,, \\
  \widehat P_{ua}(y_{ua}) &= \frac 1 {z_u} \sum_{\{y_{ub},\, b \in \dd u \sm a\}} \psi_u(y_u) \prod_{b \in \dd u \sm a} P_{bu}(y_{ub})
\end{align}
\end{subequations}
where $z_u$ and $z_a$ are normalization constants, which we shall omit in the following writing the updates up to a multiplicative constant.
The number of terms in the sums grows exponentially with the connectivity of the nodes, and we need to derive some update equations which can be computed efficiently.

Let us begin with the update equation for the users nodes. The constraint \eqref{eq:constraint_u_sum} requires that either 0 or 1 of the $y$'s can be equal to $\yS$, all the other $y$'s being equal to $\yU$ or $\yA$. Moreover, the constraint \eqref{eq:constraint_u_optimal} requires that the $y$'s associated to edges with a higher utility values than the edge with $y$ equal to $\yS$ have $y$ equal to $\yU$. We shall consider separately the three cases $y_{ua} = \yU$, $\yA$ and $\yS$. When $y_{ua} = \yU$, it is possible that all the $y$'s are $\yU$. Otherwise, exactly one of them must be equal to $\yS$, and we must sum over the possible choices for this $y$. The remaining $y$'s must be $\yU$ if they correspond to a larger value than the one with $y = \yS$ and otherwise they can be either $\yU$ or $\yA$. We obtain
\begin{subequations} \label{eq:BP_P_ua}
\begin{align}
  \widehat P_{ua}(\yU) &=\frac1{z_{ua}} \left\{\prod_{b \in \dd u \sm a} P_{bu}(\yU) + \sum_{b \in \dd u \sm a} P_{bu}(\yS) \prod_{\substack{c \in \dd u \sm a: \\ v_{uc} > v_{ub}}} P_{bu}(\yU) \prod_{\substack{d \in \dd u \sm \{a,b\}: \\ v_{ud} \leq v_{ub}}} \big[ P_{du}(\yU) + P_{du}(\yA) \big]\right\} \,.
\end{align}
When $y_{ua} = \yA$, one of the other $y$'s must be equal to $\yS$, and it must have a value at least as large as $v_{ua}$, and we obtain
\begin{align}
  \widehat P_{ua}(\yA) &=\frac1{z_{ua}} \sum_{\substack{b \in \dd u \sm a: \\ v_{ub} \geq v_{ua}}} P_{bu}(\yS) \prod_{\substack{c \in \dd u \sm a: \\ v_{uc} > v_{ub}}} P_{bu}(\yU) \prod_{\substack{d \in \dd u \sm \{a,b\}: \\ v_{ud} \leq v_{ub}}} \big[ P_{du}(\yU) + P_{du}(\yA) \big] \,.
\end{align}
Finally, when $y_{ua} = \yS$, all the other $y$'s must be either $\yU$ or $\yA$, and they can be $\yA$ only if their value is not larger than $v_{ua}$, giving
\begin{align}
  \widehat P_{ua}(\yS) &=\frac1{z_{ua}} \prod_{\substack{b \in \dd u \sm a: \\ v_{ub} > v_{ua}}} P_{bu}(\yU) \prod_{\substack{c \in \dd u \sm \{a, b\}: \\ v_{uc} \leq v_{ua}}} \big[ P_{cu}(\yU) + P_{cu}(\yA) \big] \,.
\end{align}
\end{subequations}
All these products can be computed efficiently, providing an efficient update. The normalization constant $z_{ua}$ can be fixed afterwards to ensure $ \widehat P_{ua}(\yU) +  \widehat P_{ua}(\yA)  +  \widehat P_{ua}(\yS) =1$.

In order to compute the update for service unit nodes, we introduce, for any subset $\mathsf{K} \in \dd a$ of the edges incident on service unit $a$, the convolution
\begin{align}
  \mathcal P_\mathsf{K}(S, T) &= \sum_{\{y_{ua},\, u \in \mathsf{K}\}} \1 \left[ \sum_{u \in \mathsf{K}} w_{ua} \1[y_{ua} = \yS] = T \right] \prod_{u \in \mathsf{K}} \1 \Big[ \big\{ y_{ua} = \yU \big\} \Leftrightarrow \big\{ w_{ua} + S > C_a \big\} \wedge \big\{ y_{ua} \neq \yS \big\} \Big] \widehat P_{ua}(y_{ua})
\end{align}
which can be computed efficiently thanks to the relation
\begin{align}
  \mathcal P_{\mathsf{K} \cup \mathsf{L}}(S, T) &= \sum_{T_1, T_2} \1 \big[ T_1 + T_2 = T\big] \mathcal P_\mathsf{K}(S, T_1) \mathcal P_\mathsf{L}(S, T_2)
\end{align}
valid for any disjoint subsets $\mathsf{K}$ and $\mathsf{L}$ of the incident edges, starting from the single edge quantities
\begin{align}
  \mathcal P_u(S, T) &=
  \begin{cases}
    \widehat P_{ua}(\yS) & \text{ if $T = w_{ua}$ (for any $S$)} \\
    \widehat P_{ua}(\yA) & \text{ if $T = 0$ and $S \leq C_a - w_{ua}$} \\
    \widehat P_{ua}(\yU) & \text{ if $T = 0$ and $S > C_a - w_{ua}$} \\
    0 & \text{ otherwise}
  \end{cases} & (u \in \dd a) \,.
\end{align}
In terms of $\mathcal P_{\dd a \sm u}(S, T)$ we have:
\begin{subequations} \label{eq:BP_P_au}
\begin{align}
  P_{au}(\yU) &=\frac1{z_{au}} \sum_S \1 \big[ C_a - w_{ua} < S \leq C_a \big] \mathcal P_{\dd a \sm u}(S, S) \\
  P_{au}(\yA) &=\frac1{z_{au}} \sum_S \1 \big[ 0 \leq S \leq C_a - w_{ua} \big] \mathcal P_{\dd a \sm u}(S, S) \\
  P_{au}(\yS) &=\frac1{z_{au}} \sum_S \1 \big[ 0 \leq S \leq C_a - w_{ua} \big] \mathcal P_{\dd a \sm u}(S+w_{ua}, S)
\end{align}
\end{subequations}
from which it is clear that we need to compute $\mathcal P_{\dd a \sm u}(S, T)$ for any $T \in \{0, 1, \dots, C_a\}$ and $S \in \{0, 1, \dots, C_a\}$. Again, the normalization constant $z_{au}$ can be fixed afterwards to ensure $ P_{au}(\yU)+ P_{au}(\yA)+ P_{au}(\yS)=1$.

\subsection{The ``mirror message'' approximation for the stochastic case} \label{sec:mirror}\label{SubSec:mirror}

In the stochastic case, the average value of the observables \eqref{eq:avg_O_stochastic} is computed as a double average: first over the distribution of Nash equilibria for fixed stochastic parameters $t$ (i.e. over the dynamical variables $y$), and then over the realization of $t$. In statistical physics two different kinds of averages over the stochastic parameters $t$ and the dynamical variables $y$ are considered. The first is called {\em quenched average}, and it assumes that the random parameters $t$ are kept fixed (i.e. they are ``quenched'' or ``frozen'') as the dynamical variables $y$ evolve in time. The second kind of average is called {\em annealed average}, and it assumes that the random parameters $t$ are allowed to evolve over timescales comparable with those of the dynamical variables $y$. In most cases (including ours), the correct values of the observables are provided by quenched averages, which unfortunately are difficult to compute, and annealed averages are often used as easily computed (but uncontrolled) approximations.

The distribution $P(y,t)$ corresponding to the quenched average is
\begin{align}
	P (y, t) &= P(y|t) P(t) = \frac {\mathscr C(y,t)} {\mathcal N(t)} P(t)\label{eq:mirrorexact}
\end{align}
where $\mathscr C(y,t)$ is the indicator function of the constraints \eqref{eq:constraints}, $\mathcal N(t) = \sum_y \mathscr C(y,t)$ is the number of Nash equilibria as a function of $t$, and $P(t) = \prod_u P_u(t_u)$ is the distribution of the stochastic parameters $t$. The annealed approximation of this quenched distribution is given by
\begin{align}
	P^\mathrm{ann}(y,t) = \frac 1 {Z^\mathrm{ann}} \mathscr C(y,t) P(t)\label{eq:annealed}
\end{align}
where $Z^\mathrm{ann}$ is a partition function independent of $t$. 
$P^\mathrm{ann}(y,t)$ can be viewed as an approximation to $P(y,t)$,  and it is typically a rather poor one. Perhaps the most striking evidence of their difference is that the marginal for $t$ of $P(y,t)$ is by construction the disorder distribution $P(t)$, while the marginal for $t$ of $P^\mathrm{ann}(y,t)$ will be in general different.
However, the approximation can be improved drastically in a simple and systematic way (and eventually made exact) with a method already employed by Morita \cite{morita_statistical_1964,kuhn_equilibrium_1996}, that can be reinterpreted in a natural way in terms of the cavity equations we are employing.
Formally, the quenched distribution can be rewritten as
\begin{align}
	P(y,t) &= \mathscr C(y,t) P(t) e^{\phi(t)}
\end{align}
with $\phi(t) = - \log \mathcal N(t)$. Since the $t$'s are $N$ binary variables, the function $\phi(t)$ can be parametrized uniquely as
\begin{align} \label{eq:phi_representation}
  \phi(t) &= \lambda + \sum_u \nu_u t_u + \sum_{u<v} \rho_{uv} t_u t_v + \sum_{u<v<w} \omega_{uvw} t_u t_v t_w + \dots
\end{align}
where the sum is over the $2^N$ possible subsets of users. 
By truncating the sum \eqref{eq:phi_representation}, keeping only the constant term $\lambda$, we obtain the annealed approximation \eqref{eq:annealed}, where $\lambda = -\log Z_{ann}$ is the free energy of the system. We will employ instead the next order approximation, that corresponds to keeping up to the linear terms: 
\begin{align} \label{eq:P_a_phi_1}
  P^\mathrm{lin}(y, t) &= \frac 1 {Z^\mathrm{lin}} \prod_{u \in \mathsf{U}} \mathscr C_u(y_u, t_u) P_u(t_u) e^{\nu_u t_u} \prod_{a \in \mathsf{S}} \mathscr C_a(y_a).
\end{align}

The value $\nu_u$ can be thought of as a parameter of a prior distribution of $t_u$; but its value is not really known {\em a priori} and will be found self-consistently. Note that \eqref{eq:P_a_phi_1} has the same factorization we had in the deterministic case, and which can be represented by the factor graph of Figure \ref{fig:factor_graphs} (right-hand panel), that has the same topology as the graph $\mathsf{G} = (\mathsf{U}, \mathsf{S}; \mathsf{E})$ representing the instance. 
In order to determine coefficients $\nu_u$, we will impose the constraints the average value of $t_u$ is the same as in the exact expression \eqref{eq:mirrorexact}, i.e. the distribution of our disorder variables $P(t_u)$. 
The marginal of $t_u$ is proportional to $Q_u(t_u) \widehat Q_u(t_u)$, where the message $\widehat Q_u(t_u)$ is determined by the constraint $\mathscr C_u(y_u, t_u)$ and by the messages entering the factor node representing it, and we obtain
\begin{align}
  Q_u(0) &= \frac{ P_u(0) / \widehat Q_u(0) } { P_u(0) / \widehat Q_u(0) + P_u(1) / \widehat Q_u(1) } \,, & 
  Q_u(1) &= \frac{ P_u(1) / \widehat Q_u(1) } { P_u(0) / \widehat Q_u(0) + P_u(1) / \widehat Q_u(1) } \,.
\end{align}
These equations have a simple intuitive interpretation: the effect of the message $Q_u \propto P_u /\widehat Q_u  $ is to counterbalance the bias on $t_u$ induced by $\widehat Q_u$, i.e. by the rest of the system, to restore the correct marginal $P_u$. We therefore refer to the factor node which determines $Q_u$ as a \emph{mirror} node, and to $Q_u$ as a \emph{mirror message}.

Higher order terms in \eqref{eq:phi_representation} could in principle be retained. For instance, for the next order, one can expect that the correlation between $t_u$ and $t_v$ will be strongest if $u$ and $v$ are close in the graph. One could use a generalized Belief Propagation scheme \cite{yedidia_generalized_2000, yedidia_understanding_2003} to perform the computation by defining appropriate regions on the factor graph (at the cost of a larger computational effort), and imposing that pairwise correlations are identical to the ones of the distribution of the quenched disorder (in our case, connected correlations are zero). In the following we shall only consider the linear approximation \eqref{eq:P_a_phi_1} because, as we shall verify ex-post, the results it gives are sufficiently accurate.

The BP equation \eqref{eq:BP_P_au} for $P_{au}(y_{ua})$ remains unchanged, while the BP equation \eqref{eq:BP_P_ua} remains the same only if $t_u = 1$. If $t_u = 0$ the constraint on the local configuration of $y_u$ is that $y_{ua}$ can be either 0 or 1 for all $a \in \dd u$. We therefore obtain:
\begin{subequations} \label{eq:BP_P_stoch}
\begin{align}
  \widehat P_{ua}(\yU) &\propto Q_u(0) \prod_{b \in \dd u \sm a} \big[ P_{bu}(\yU) + P_{bu}(\yA) \big] + \nonumber \\
  &\quad + Q_u(1) \Bigg\{ \prod_{b \in \dd u \sm a} P_{bu}(\yU)+ \sum_{b \in \dd u \sm a} P_{bu}(\yS) \prod_{\substack{c \in \dd u \sm a: \\ v_{uc} > v_{ub}}} P_{bu}(\yU) \prod_{\substack{d \in \dd u \sm \{a,b\}: \\ v_{ud} \leq v_{ub}}} \big[ P_{du}(\yU) + P_{du}(\yA) \big] \Bigg\} \,, \\
  \widehat P_{ua}(\yA) &\propto Q_u(0) \prod_{b \in \dd u \sm a} \big[ P_{bu}(\yU) + P_{bu}(\yA) \big] + \nonumber \\
  &\quad + Q_u(1) \sum_{\substack{b \in \dd u \sm a: \\ v_{ub} \geq v_{ua}}} P_{bu}(\yS) \prod_{\substack{c \in \dd u \sm a: \\ v_{uc} > v_{ub}}} P_{bu}(\yU) \prod_{\substack{d \in \dd u \sm \{a,b\}: \\ v_{ud} \leq v_{ub}}} \big[ P_{du}(\yU) + P_{du}(\yA) \big] \,, \\
  \widehat P_{ua}(\yS) &\propto Q_u(1) \prod_{\substack{b \in \dd u \sm a: \\ v_{ub} > v_{ua}}} P_{bu}(\yU) \prod_{\substack{c \in \dd u \sm \{a, b\}: \\ v_{uc} \leq v_{ua}}} \big[ P_{cu}(\yU) + P_{cu}(\yA) \big] \,.
\end{align}
\end{subequations}
These updates can be computed as efficiently as their deterministic counterparts \eqref{eq:BP_P_ua}.
One more message exits the factor node $u$: the message to the variable $t_u$, which we denote by $\widehat Q_u(t_u)$. The corresponding update equation is easily seen to be
\begin{align} \label{eq:BP_nu_hat}
  \widehat Q_u(0) &\propto \prod_{a \in \dd u} \big[ P_{au}(\yU) + P_{au}(\yA) \big] \,, \\
  \widehat Q_u(1) &\propto \prod_{a \in \dd u} P_{au}(\yU)+ \sum_{a \in \dd u} P_{au}(\yS) \prod_{\substack{b \in \dd u: \\ v_{ub} > v_{ua}}} P_{au}(\yU) \prod_{\substack{c \in \dd u \sm a: \\ v_{uc} \leq v_{ua}}} \big[ P_{cu}(\yU) + P_{cu}(\yA) \big] \,.
\end{align}
Finally, the update equation for the fields $Q_u(t)$ are obtained by requiring that the marginal for $t_u$ is equal to $P_u(t_u) = p_u t_u + (1-p_u) (1-t_u)$:
\begin{align}
  Q_u(0) \widehat Q_u(0) &\propto 1-p_u & Q_u(1) \widehat Q_u(1) &\propto p_u
\end{align}
from which we obtain
\begin{subequations} \label {eq:BP_nu}
\begin{align}
  Q_u(0) &\propto \frac {(1-p_u) \widehat Q_u(1)} {p_u \widehat Q_u(0) + (1-p_u) \widehat Q_u(1)} \,, \\
  Q_u(1) &\propto \frac {p_u \widehat Q_u(0)} {p_u \widehat Q_u(0) + (1-p_u) \widehat Q_u(1)} \,.
\end{align}
\end{subequations}

\setcounter{subsection}{0}
\section*{Results}
\subsection*{Numerical results on a random ensemble of deterministic instances} 

\subsection{Results on deterministic instances} \label{SubSec:ensemble_det}\label{sec:numerical_results}\label{SubSec:dynamics} \label{SubSec:entropy}\label{SubSec:generalNE} 
\subsubsection*{Definition of the ensemble} 
We consider a random ensemble of deterministic instances with $N$ users and $M$ service units, all with capacity $C$. For any user $u$ and any service unit $a$, the edge $(ua)$ is present with probability $q$, and the parameters $w_{ua}$ and $v_{ua}$ are integers extracted from the maximum entropy distribution over the range $\{w_\mathrm{min}, \dots, w_\mathrm{max}\} \times \{ v_\mathrm{min}, \dots, v_\mathrm{max} \}$ conditioned on a given value of Pearson's correlation coefficient
\begin{align}
  c = \frac{ \left< w_{ua} v_{ua} \right> - \left< w_{ua} \right> \left< v_{ua} \right> } { \sqrt{ \left< w_{ua}^2 \right> - \left< w_{ua} \right>^2} \sqrt{ \left< v_{ua}^2 \right> - \left< v_{ua} \right>^2} }
\end{align}
(the reason for this choice will be clear shortly). Such an ensemble is fully specified by the parameters $N$, $M$, $C$, $q$, $w_\mathrm{min}$, $w_\mathrm{max}$, $v_\mathrm{min}$, $v_\mathrm{max}$ and $c$. These parameters will determine the qualitative features of the phenomenology as follows.

The total available capacity $\hat{C} = MC$ can be compared to a lower and an upper bound for the total capacity required to service all users, defined as
\begin{align} \label{eq:C_bounds}
  \hat{C}^- &= \sum_u \min_{a \in \dd u} w_{ua} \,, & \hat{C}^+ &= \sum_u \max_{a \in \dd u} w_{ua} \,.
\end{align}
When the total available capacity $\hat{C}$ is large compared to $\hat{C}^+$ the system is under-constrained and the solution is trivial: every user can be served by the service unit with the highest utility, and every service unit has some spare capacity, so that suboptimal configurations in which some users are not fully satisfied are not Nash equilibria. At the opposite end of the spectrum, when $\hat{C}$ is small compared to $\hat{C}^-$, many users receive no service at all, and the users who are served typically enjoy a low utility. As $\hat{C}$ goes to zero the number of equilibria decreases, and it reaches 1 when $\hat{C}$ is smaller than the smallest weight and all the users are unserved, which is again a trivial regime. The interesting regime corresponds to intermediate values $\hat{C}^- \lesssim \hat{C} \lesssim \hat{C}^+$, when some of the capacity constraints are saturated and some other are not.

For any value of the total capacity $\hat{C}$, the level of tightness of the constraints depends on two more parameters. The first is the average connectivity of users, determined by $q$: for larger values of $q$, users will have more alternative service units to chose from, and the system will be less constrained. The second is the minimum value of the weights $w_\mathrm{min}$: spare capacity on individual service units smaller than $w_\mathrm{min}$ cannot be used, so that for larger values of $w_\mathrm{min}$ typical configurations will be more inefficient and the system will be more constrained.

Finally, the correlation $c$ between weight and utility is a measure of the degree of competition among users: when $c$ is close to $-1$, users prefer to be served by those service units over which they place a low burden, minimizing the capacity they subtract to other users, and the competition is mild; when $c$ is close to $0$, users' preferences are independent of the load they place on service units, and on the impact this has on the capacity available to other users; finally, when $c$ is close to $+1$, weights and utility values tend to coincide (up to an affinity transformation), and users try to subtract as much capacity as possible to other users, so that the competition is harsh.

\subsubsection*{Average values of the observables: $\mathcal U$, $\mathcal D$ and $\mathcal C^*$}

\begin{figure}[t]
\begin{center}
   \includegraphics[width=0.9\columnwidth]{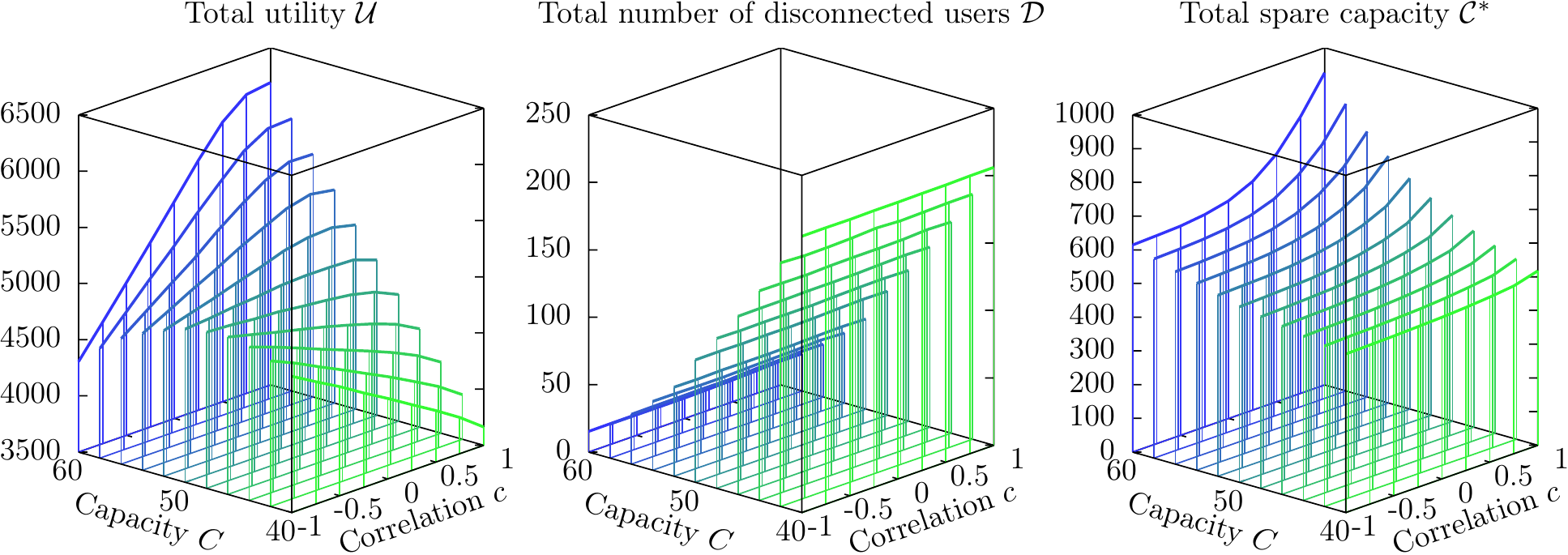}
  \caption{\label{fig:obs_vs_C_and_c} Average values $\mathcal U$, $\mathcal D$ and $\mathcal C^*$ of the observables as a function of the capacity $C_a$ of service units and the correlation $c$ between weight and utility on individual edges. The other parameters are $N=1\,000$, $M=200$, $q=0.04$, $w_\mathrm{min}=6$, $w_\mathrm{max}=15$, $v_\mathrm{min}=1$ and $v_\mathrm{max}=10$. Each data point, corresponding to a vertical line, is an average over $115$ instances, and the standard deviations are of the order of the width of the lines.}
  \end{center}
\end{figure}

We study the average values of the relevant observables (i.e. the total utility, the total number of disconnected users, and the total spare capacity) as a function of the capacity $C$ of individual service units and of the correlation $c$ between the weight $w_{ua}$ and the values $v_{ua}$ on each edge $(ua) \in \mathsf{E}$, keeping fixed the remaining parameters. The range of values for $C$ corresponds to a total capacity $\hat{C}$ between $8\,000$ and $12\,000$, spanning the relevant range of values defined by the bounds \eqref{eq:C_bounds} with expected values $\hat{C}^- = 6\,581$ and $\hat{C}^+ = 14\,418$. Results of numerical simulations are shown in Figure \ref{fig:obs_vs_C_and_c}.

For small values of $C$, the total utility $\mathcal U$ decreases with $c$ as expected. However, for larger values there is a surprising inversion of this dependency: larger values of $c$, which give rise to harsher competition between users, correspond to \emph{higher} values of the average total utility. In particular, when the individual capacity is $C = 60$, the average total utility for $c = -1$ is $4\,306 \pm 12$, while for $c = +1$ it is $6\,190 \pm 9$, with a $43.8\%$ increase. It is also surprising that, for values of $c$ close to $-1$, the average total utility \emph{does not} increase monotonically with the capacity $C$: on the contrary, for $c = -1$ it reaches a maximum value $4\,896 \pm 35$ for $C = 47$ (not plotted in Figure \ref{fig:obs_vs_C_and_c}), well above the value corresponding to $C=60$, which is $4\,306 \pm 12$.

The average total number of disconnected users $\mathcal D$ shows, as expected, a strong dependency on the capacity $C$, but surprisingly it is almost independent on the correlation $c$. Since the social welfare depends on both $\mathcal U$ and $\mathcal D$, we obtain the striking result that, provided the capacity $C$ is large enough, a harsher competition between users, in which each user prefers to subtract to other users as many of the available resources as possible, gives rise on average to a higher social welfare than a milder form of competition, in which users prefer to minimize the amount of resources they subtract to other users.

Finally, the average total spare capacity $\mathcal C^*$ increases monotonically with $C$, which could be easily expected, and also with $c$: as the average weight placed on service units by individual users increases, it becomes more difficult to use the capacity efficiently, and a larger fraction of capacity is wasted (even though a sizable number of users are disconnected!).

\subsubsection*{Comparison with explicit dynamics} 

\begin{figure}[t]
\begin{center}
  \includegraphics[width=0.9\columnwidth]{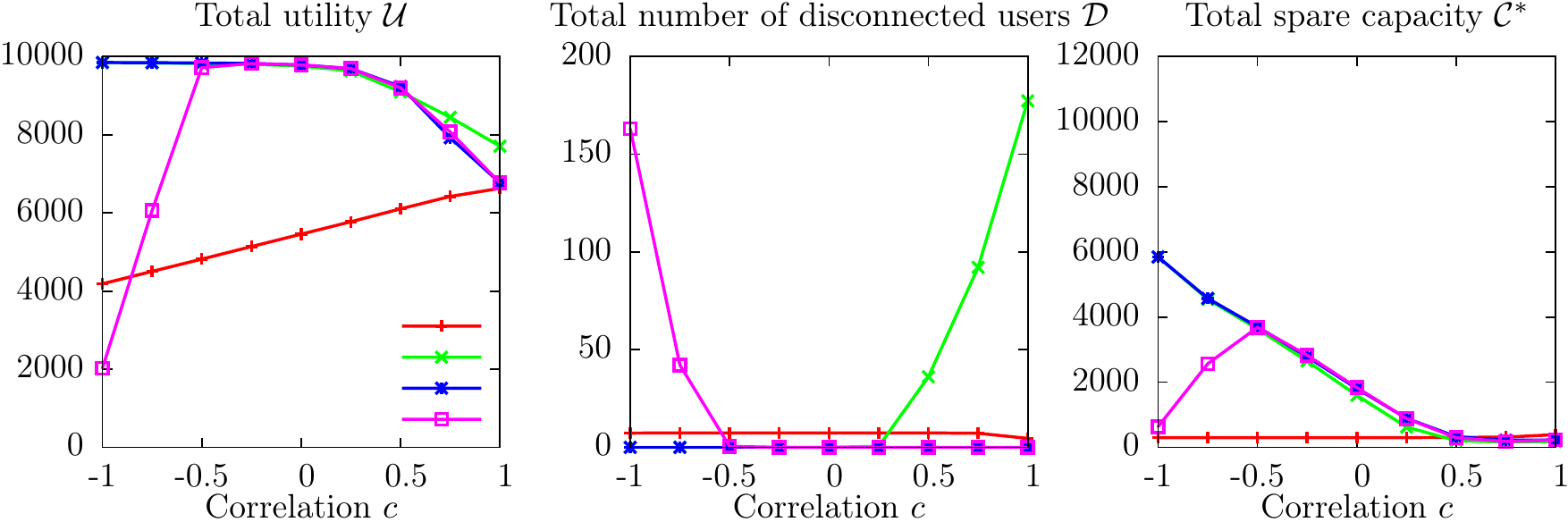}
  \caption{\label{fig:UDC_vs_c} Average values $\mathcal U$, $\mathcal D$ and $\mathcal C^*$ of the observables as a function of the correlation $c$ between weight and utility on individual edges. The other parameters are $N=1\,000$, $M=100$, $C=120$, $q=0.2$, $w_\mathrm{min} = 6$, $w_\mathrm{max} = 15$, $v_\mathrm{min} = 1$ and $v_\mathrm{max} = 10$. Each data point is an average over $40$ instances, and for each instance, each dynamics is realized $10\,000$ times. The average value of each observable is computed over the instances and over the realizations. The standard deviations (of the averages over the realizations of the dynamics across different instances) are much smaller than symbol sizes.}
  \end{center}
\end{figure}

We compare the average values of the observables obtained by averaging uniformly over all Nash equilibria with Belief Propagation with the results of the explicit simulation of three interesting dynamics which converge to Nash equilibria:
\begin{itemize}
  \item \emph{Greedy (G)} -- The first dynamics consists in extracting a random permutation of the users and assigning to each one in turn the service unit with the highest utility among the available ones. Obviously, at the end of the assignment we have a Nash equilibrium, in which some users (who came early in the permutation) enjoy a very high utility while some other users (who came late) are either disconnected or with very low utility.
  \item \emph{Best response (BR)} -- The second dynamics starts from a random initial condition, extracted by forming again a random permutation of the users and assigning to each one in turn a service unit extracted uniformly at random among the available ones. The dynamics then proceeds in rounds. In each round, a random permutation of the users is extracted, and each user in turn attempts to improve their utility (possibly freeing some capacity at the service unit previously serving them, and allowing other users to use it). The dynamics stops when no user can improve their utility (and therefore the configuration is a Nash equilibrium).
  \item \emph{Best response from ``bad'' initial condition (BRB)} -- The third dynamics is identical to the second one, except for the choice of the initial condition: instead of selecting uniformly at random their service unit among the available ones, each user initially selects the \emph{worst} one among the available ones (i.e. the one with the lowest utility), in turn and according to a random permutation of the users. Then, they follow the best response dynamics.
\end{itemize}
Numerical results for the average values of the observables $\mathcal U$, $\mathcal D$ and $\mathcal C^*$ as a function of the correlation $c$ (for fixed values of the remaining parameters) are shown in Figure \ref{fig:UDC_vs_c} for BP and for the three dynamics (see caption for simulation details).

The most striking feature of these plots is that the uniform average over Nash equilibria computed by BP is very far from the averages computed by sampling over the three dynamics, even in the region where the three dynamics give very similar results (i.e. for $-0.5 \leq c \leq 0.5$). It appears very clearly that the three dynamics we considered are strongly biased towards ``good'' equilibria. In particular, when $-0.5 \leq c \leq 0.25$ all the dynamics converge to equilibria which are nearly optimal, with values of $\mathcal U$ very close to the expected value $\left< U^+ \right> \simeq 9\,843$ of the theoretical upper bound $U^+ =  \sum_u \max_{a \in \dd u} v_{ua}$, while BP finds values which grow almost linearly with $c$ (again, indicating that harsher competition leads on average to higher total utility) in the range from $4\,817 \pm 22$ (where the error is the standard deviation of the average of the observable across different instances) obtained for $c = -0.5$ to $5\,775 \pm 19$ obtained for $c = 0.25$, which is between $41.3\%$ and $51.8\%$ less than the optimum.

When $c = -1$, BRB actually finds equilibria with much lower average total utility than BP: in the initial condition of BRB users are connected to service units with low utility values, which for $c = -1$ correspond to high weights, so that service units are saturated (as confirmed by the low value of the average total spare capacity $\mathcal C^* = 635 \pm 70$), and the configuration is close to an equilibrium, which the best response part of the dynamics reaches without improving much the value of $\mathcal U$. The average total utility of BRB increases very sharply as $c$ passes from $-1$ to $-0.5$, where it is already very close to the optimum, indicating that when the anticorrelation between weights and values is imperfect, the initial configuration is no longer close to an equilibrium (as confirmed by the much larger values of $\mathcal C^*$), and during the best response part of the dynamics the configuration can ``escape'' and reach a good equilibrium. As $c$ increases further, all the dynamics find near optimal equilibria up to $c = 0.25$ and then the value of $\mathcal U$ starts to decrease significantly (as one would normally expect). When $c = 1$ BP finds results that are similar to the two best response dynamics, but still far from greedy. In the full range of $c$ the average total utility found by BP increases almost linearly with $c$.

The results found by different dynamics for the average total number of disconnected users $\mathcal D$ is even more heterogeneous. With BR, all users are \emph{always} connected provided $c < 0.75$, and for larger values of $c$ the value of $\mathcal D$ is never larger than $6.2 \times 10^{-3}$. This is in sharp contrast with the results of BRB, which finds a relatively high number of disconnected users when $c$ is small (with $\mathcal D = 163 \pm 3$ for $c = -1$), and with the results of greedy, which finds a relatively large number of disconnected users when $c$ is large (with $\mathcal D = 177 \pm 2$ for $c = 1$). By contrast, BP finds a value of $\mathcal D = 7.4 \pm 0.2$ which is constant for $-1 \leq c \leq 0.5$, and then decreases to $\mathcal D = 4.6 \pm 0.2$ for $c = 1$. In the region $-0.5 \leq c \leq 0.25$ the three dynamics find negligible values of $\mathcal D$, again indicating a strong bias towards high utilities in the equilibria they reach.

Finally, the average value of the total spare capacity $\mathcal C^*$ found by BP is almost constant and very low across the range of values of $c$, with $\mathcal C^* = 302 \pm 1$ for $c \leq 0.5$ and increasing slightly for larger values of $c$ to $393 \pm 2$ for $c = 1$. It appears very clearly that the uniform average over Nash equilibria is dominated by configurations with utilities that are much smaller than the optimal one, but which are ``locked'' by a lack of spare capacity. The same thing seems to happen to BRB for $c \leq -0.75$, and to other dynamics (but to a much lesser extent) for $c \geq 0.5$.

\subsubsection*{Analysis of the entropy vs. utility curve for individual instances}

\begin{figure}[h]
\begin{center}
    \includegraphics[width=0.8\columnwidth]{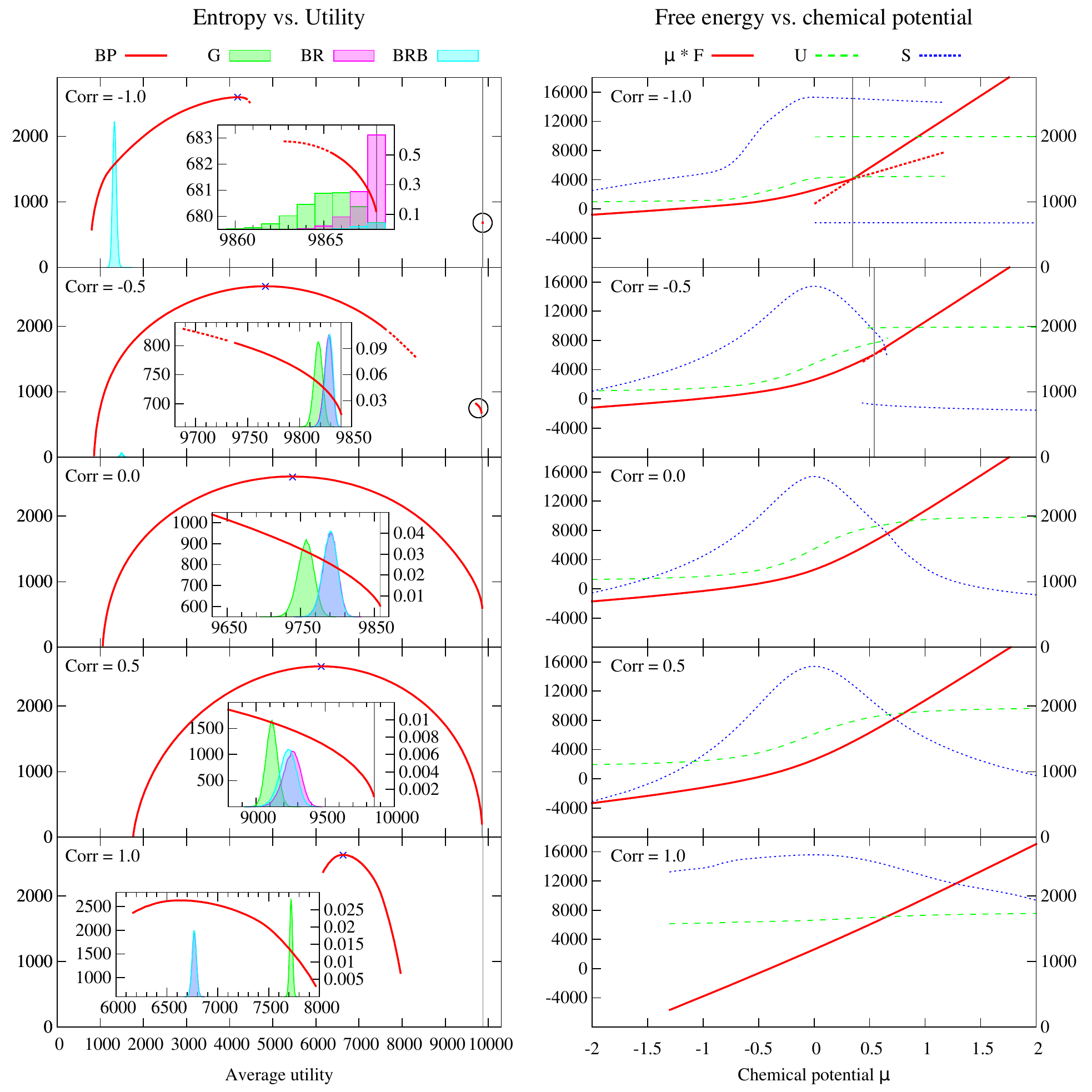}
  \caption{\label{fig:entropy} Computation of the average utility $\mathcal U$ and of the entropy $S$ as a function of the parameter $\mu$ for single instances with correlation $c$ equal to $-1$, $-0.5$, $0$, $0.5$ and $1$ (from top to bottom). The other parameters are $N=1\,000$, $M=100$, $C=120$, $q=0.2$, $w_\mathrm{min}=6$, $w_\mathrm{max}=15$, $v_\mathrm{min}=1$ and $v_\mathrm{max}=10$. The values of $\mathcal U$ and $S$ vs. $\mu$ are shown on the right-hand column of plots, together with the Bethe free energy $\mu F = \mu \, \mathcal U + S$. The values of $\mathcal U$ and $\mu F$ are read on the left-hand scale, the values of $S$ on the right-hand one. For $c\leq -0.5$ a first order transition, corresponding to a discontinuity in the first derivative of the free energy, is clearly visible. The critical value $\mu^*$ is marked by a vertical grey line, and the thermodynamically unstable branches of the free energy are dotted. The entropy is plotted as a function of the average utility on the left-hand column of plots. For $c \leq -0.5$ the entropy curve has two branches, separated by a wide gap (the high utility branches, which cover a small interval of utilities, are shown in detail in the insets, and they are highlighted by a black circle in the main plots). Dotted lines correspond to the thermodynamically unstable branches of the free energy. For $c \geq 0$ the entropy curve is uninterrupted. The blue crosses at the top of each curve represent the values obtained with $\mu = 0$, i.e. the uniform average over all Nash equilibria, and they always coincide with the maximum of the entropy. The histograms represent the distribution of the total utilities found by the dynamics G, BR and BRB. The number of runs for each dynamics is always $10^5$, except for BRB with $c = -1$ where it is $10^6$. The frequency of each value of utility can be read on the right-hand scale of the insets. The gray vertical lines are upper bounds to the total utility defined as $U^+ = \sum_u \max_{a \in \dd u} v_{ua}$.}
\end{center}
\end{figure}

In order clarify the (sometimes counterintuitive) phenomena discussed in the previous paragraphs, we analyzed in detail individual instances with different values of the correlation $c \in \{-1, -0.5, 0, 0.5, 1\}$ (with the other parameters taking the same values as in the previous paragraph). Specifically, we computed with BP the thermodynamic entropy $S(\mu)$ and the average total utility $\mathcal U(\mu)$ for a large number of values of the parameter $\mu$, both positive and negative. This allows to plot $S(\mu)$ vs. $\mathcal U(\mu)$, providing a measure of the number of Nash equilibria corresponding to a given value of the total utility. We compared this to a sampling of the equilibria obtained with the three dynamics previously introduced: greedy (G), best response from random initial condition (BR) and best response from ``bad'' initial conditions (BRB). The results are shown in Figure \ref{fig:entropy}.

For $c = -1$ a first order transition is present: the free energy  $\mu F(\mu) = \mu \, \mathcal U(\mu) + S(\mu)$ has a discontinuous derivative at $\mu^* = 0.3453$, and the entropy vs. utility curve consists of two branches separated by a wide gap. The thermodynamically stable branch is the low-utility one for $\mu < \mu^*$ and the high-utility one for $\mu > \mu^*$. 
In the high utility branch, both $\mathcal U(\mu)$ and $S(\mu)$ converge to a constant limit as $\mu \to +\infty$, varying very little between $\mu = 1$ (with $\mathcal U = 9\,867.23$ and $S = 681.33$) and $\mu = 10/3$ (with $\mathcal U = 9\,867.97$ and $S = 680.20$). The limit value for $\mathcal U$ coincides with the upperbound for the utility $U^+ = \sum_u \max_{a \in \dd u} v_{ua} = 9\,868$, and the limit value of $S$ gives the logarithm of the number of Nash equilibria with $\mathcal U = U^+$, indicating that this upperbound is feasible and that the number of optimal equilibria is exponentially large. In fact, the three dynamics G, BR and BRB are all capable of finding equilibria with $\mathcal U = U^+$. We also found optimal equilibria with $\mathcal U = U^+$ with the reinforced BP \cite{braunstein_learning_2006,baldassi_theory_2013} or Max-Sum algorithms \cite{altarelli_statistical_2009,altarelli_optimizing_2013}, the zero temperature (or infinite $\mu$, in this case) version of BP (see Table \ref{tab:BP_vs_MS} for the full results of Max-Sum). For $\mu > 10/3$, BP converges to a fixed point with unphysical properties ($\mathcal U > U^+$ and $S < 0$). The high utility branch continues for $\mu < \mu^*$, and it is possible to explore it with BP starting with $\mu > \mu^\mathrm{BP} = 0.7812$ and decreasing it in small steps, allowing BP to converge between each step (and keeping the messages when $\mu$ changes). Notice that for $\mu^* < \mu < \mu^\mathrm{BP}$ the stable branch is the high utility one, but BP converges to (unstable) solutions in the low utility branch. The values of $\mathcal U$ and $S$ tend to a constant limit as $\mu \to 0^+$, with $\mathcal U = 9\,863.67$ and $S = 682.82$ for $\mu = 0.1$ which become $\mathcal U = 9\,862.67$ and $S = 682.87$ for $\mu = 10^{-6}$. The results obtained with greedy show that Nash equilibria can be found with utilities as small as  $9\,857$, which is smaller than the limit value found by BP.

\begin{table}
\begin{center}
  \setlength{\tabcolsep}{6pt}
  \begin{tabular}{|c|cc|ccc|c|}
    \hline
    $c$ & BP & MS & BP & MS & $ U^+$ & PoA \\
    \hline
    $-1$ & $800.00$ & $800$ & $9\,867.97$ & $9\,868$ & $9\,868$ & $12.34$\\
    $-0.5$ & $856.89$ & $857$ & $9\,839.98$ & $9\,840$ & $9\,840$ & $11.48$ \\
    $0$ & $1\,064.44$ & $1\,057$ & $9\,837.02$ & $9\,858$ & $9\,858$ & $9.33$ \\
    $0.5$ & $1\,763.94$ & $1\,782$ & $9\,851.02$ & $9\,853$ & $9\,855$ & $5.53$ \\
    $1$ &  $6\,169.17$ & $6\,139$ & $7\,962.71$ & $7\,965$ & $9\,874$ & $1.30$ \\
    \hline
  \end{tabular}
  \caption{\label{tab:BP_vs_MS} Comparison of the minimum and maximum utilities found by Belief Propagation as $\mu \to \pm \infty$ with the values found by Max-Sum, which allows to explicitly find an equilibrium configuration of extreme utility, and with the upper bound $U^+$ defined in the main text. The last column shows the price of anarchy computed from the MS values.}
  \end{center}
\end{table}

The low utility branch has much larger values of the entropy, and therefore it dominates the statistics. When $\mu = 0$ the distribution is unbiased (i.e. uniform over all Nash equilibria) and $\mathcal U = 4\,182.40$ is the average utility over all Nash equilibria while $S = 2\,598.23$ is the logarithm of the total number of equilibria. When $\mu^* < \mu$ the low utility branch is unstable, but it can be studied with BP as explained above, starting with a small value of $\mu$ and increasing it gradually up to $\mu = \mu^\mathrm{BP}$, where the solution found by BP jumps discontinuously to the high utility branch. For negative values of $\mu$, the distribution is biased towards Nash equilibria with lower-than-average utilities. As $\mu \to -\infty$ both $\mathcal U$ and $S$ tend to a constant limit, with $\mathcal U = 800.16$ and $S = 580.09$ at $\mu = -10$ which become $\mathcal U = 800.00$ and $S = 578.34$ at $\mu = -200/3$, after which BP starts to converge to an unphysical fixed point with zero utility and negative entropy. The average spare capacity for $\mu \leq -10$ is exactly zero, indicating that an exponential number of configurations exist with $\mathcal U = 800$ and which use all the available capacity. In fact, when $c = -1$ the edges with the lowest utility, $v_\mathrm{min} = 1$, also have the highest weight, $w_\mathrm{max} = 15$, so that if $800$ users are using edges with $v = 1$ and $w = 15$ the total utility will be $800$ and the total capacity used will be $12\,000$, with no spare capacity. Such a configuration can be easily found with Max-Sum, and we have verified that it exists. Equilibria with very low utilities can also be found by dynamics: in $93.2\%$ of $10^6$ runs BRB converges to equilibria with utilities between $1\,135$ and $1\,734$. In the remaining cases it manages to ``escape'' from the low utility branch and to reach optimal or nearly optimal equilibria.

The most striking feature of the $S$ vs. $\mathcal U$ curve for $c = -1$ is the presence of a very wide gap, covering the range $\mathcal U \in [4\,463.88,\,9\,862.78]$. This gap can be intuitively explained with the following argument. When $c = -1$, two type of equilibria exist: in ``good'' equilibria, users are served by service units with high utility, and therefore low weight, which ensures that the capacity is sufficient for (almost) all users; in ``bad'' equilibria, users are served by service units with low utility, and therefore high weight, so that there is no spare capacity to permit a dynamical transition to a good equilibrium. The maximum spare capacity in bad equilibria is approximately $M (w_\mathrm{min} - 1)$, which in our case is 500, because if the spare capacity were larger than that, at least one of the service units would have enough spare capacity to serve a user with the minimum weight, i.e. with the maximum utility, and this user would switch to it. (The spare capacity can be larger than 500 only if there are some service units connected only to edges with weight larger than $w_\mathrm{min}$, which is very unlikely in our case given that the average connectivity of service units is 200, and that the probability that an edge has minimum weight is $0.1$). If all the users are served, the total load  $L$, that is the sum of the used weights, is related to the total utility $U$ by the simple relation $U + L = N(v_\mathrm{min} + w_\mathrm{max}) = 16\,000$, so the maximum utility corresponds to the minimum weight (i.e. $11\,500$, given that the spare capacity cannot be larger than 500) which gives $4\,500$. If instead there are $D$ users who are disconnected (i.e. who are not being served), the relation between $U$ and $L$ becomes $U + L = (N - D) (v_\mathrm{min} + w_\mathrm{max})$, and the utility can only decrease. So, in bad equilibria utilities must be smaller than $4\,500$. On the other hand, in optimal equilibria, nearly all users have the maximum utility $v_\mathrm{max} = 10$, which corresponds to the minimum weight $w_\mathrm{min} = 6$, so the total weight is approximately $6\,000$, which is half the available capacity. In order for an equilibrium to be good but not optimal, some service unit $s$ must be saturated, so that its spare capacity is smaller than $w_\mathrm{min}$, and some of the users who would prefer to be served by $s$ are forced to accept a lower utility with some other service unit. Even in the extreme case in which half the service units are saturated and the other half are empty, it is very likely that a frustrated user will be able to be served by a service unit with $v_\mathrm{max} - 1$, so the maximum utility loss (relative to the optimum) is of the order of $1\,000$. This gives a lower bound for the utility in good equilibria which is approximately $9\,000$. Therefore, in the range of utilities between (approximately) $4\,500$ and $9\,000$, we expect to find no equilibria.

For $c = -0.5$ the phenomenology and its interpretation are very similar to the case $c = -1$ discussed above. However, the size of the gap in utilities is much reduced (it now covers the range $[7\,621.00,\, 9\,687.20]$), the fraction of runs of BRB which find equilibria with low utilities is much smaller ($2.0\%$), and the entropy corresponding to the minimum utility is now zero, indicating that the number of minima is subexponential. For large utilities the entropy curve ends at $\mathcal U = U^+ = 9\,840$ with a large value ($S = 682.16$), and both BR and BRB succeed in finding optimal equilibria.
For $c = 0$ and $c = 0.5$ the gap in utilities disappears, and all of the three dynamics we tested always find equilibria with large utilities, but they never achieve the upper bound $U^+$, which coincides with the end of the entropy curve (with finite values of $S$).
Finally, for $c = 1$ there is one more significant change in the phenomenology: the entropy curve now covers a much smaller range of utilities, from $6\,169.17$ to $7\,962.71$. In particular, optimal equilibria (in which every user enjoys the maximum utility) seem no longer achievable. All the three dynamics find equilibria with utilities in a narrow range, which is very different for G compared to the two best response dynamics (BR and BRB), but which fall in the range of utilities found by BP. We can estimate the range of utilities of equilibria with a simple mean-field argument. For $c = 1$, and with $w_\mathrm{max} - w_\mathrm{min} = v_\mathrm{max} - v_\mathrm{min}$, the relation between utility and weight on any edge is $w = v + \delta$, where $\delta = w_\mathrm{max} - v_\mathrm{max}$. If we denote by $\overline v$ the average value of the service utilities and by $D$ the number of disconnected users, the total utility and weight will be given by $U = (N-D) \overline v$ and $W = (N-D)(\overline v + \delta)$. The maximum utility is obtained by maximizing $U$ with respect to $\overline v$ and $D$ subject to the capacity constraint $L \leq M C$, which gives $U = 8\,000$ with $D=0$ and $\overline v=8$. The minimum utility is obtained by minimizing $U$ subject to the constraint that the average spare capacity does not exceed the average weight (because otherwise the ``average user'' could improve their utility by switching a service unit with excessive spare capacity), i.e. that $C - L/M \leq \overline v + \delta$, which gives $U = 65\,000/11 \simeq 5\,909$ with $D=0$ and $\overline v = 65/11$. The extrema of the range of values found by BP are very close to these bounds.

In summary, this analysis of individual instances allows us to understand in detail the phenomenology we observed, and to draw some general conclusions. First, we find that each one of the three dynamics we tested samples equilibria within a very narrow range of utilities (or possibly two very narrow ranges of utilities, for BRB at $c \leq -0.5$), while the full range of possible equilibria is extremely wide. Second, we clarify the reason for the increase of the average total utility when $c$ increases (i.e. when the competition becomes harsher): in the most numerous equilibria most service units are saturated, which requires the total load $L$ to be close to the capacity, i.e. the maximum possible value for $L$. As the correlation between weight and utility increases, it becomes more and more difficult to find equilibria with very low utilities, and the average utility increases. And third, our characterization also allows us to estimate the price of anarchy (i.e. the ratio between the utilities of the social optimum, which in the service provision game is always a Nash equilibrium itself, and of the worst possible equilibrium), which decreases smoothly from $12.34$ for $c = -1$ to $1.30$ for $c = 1$ (see \ref{tab:BP_vs_MS}).
 
\subsubsection*{Finding Nash equilibria with general values of utility.}
The results presented in the last two sections have shown that the best-response dynamics tend to converge towards Nash equilibria with large aggregate utility even when the initial condition is a configuration with the worst possible values of utility (see BRB results in Fig.\ref{fig:UDC_vs_c}), unless the dynamical process gets stuck because of capacity constraints. This is possibly due to the fact that the aggregate utility is a potential function for the game that always increases during the best response dynamics. If the available capacity is large, the rearrangement path due to best response  (usually called ``improvement path'' in potential games \cite{monderer_potential_1996}) is long and reaches Nash equilibria with very high utility. In a low capacity regime, instead, the improvement paths is much shorter and the best response dynamics converge after very little utility gain. If so, it should be possible to get stuck at any value of the aggregate utility in the interval of existence indicated by the BP analysis.  

\begin{figure}[h]
\begin{center}
   \includegraphics[width=0.6\columnwidth]{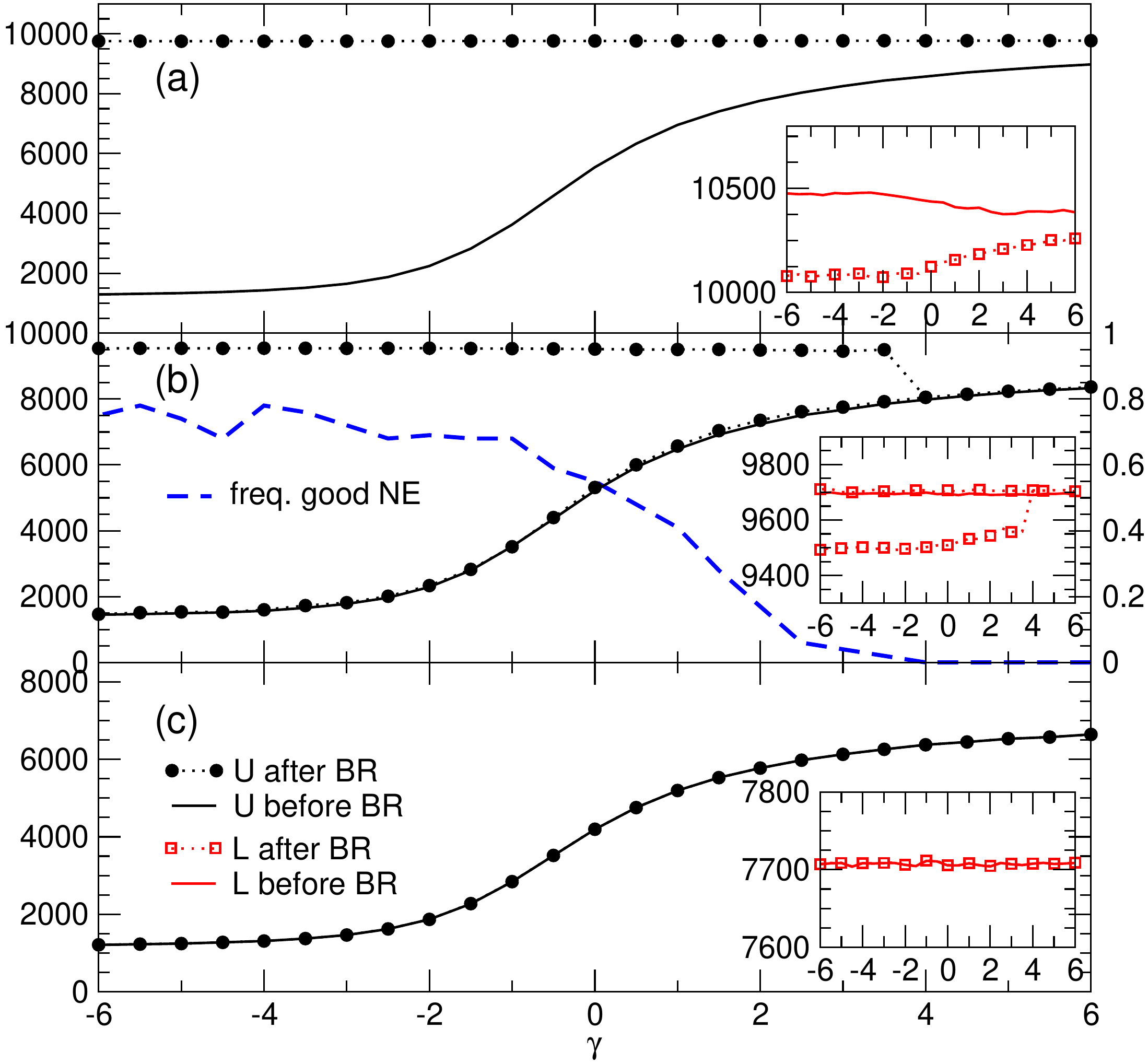}
  \caption{\label{fig:gammaBR} Average aggregate utility of the initial configuration (black full line) and of the Nash equilibria found by best response (black circles) as a function of the parameter $\gamma$ and for different values of the capacity $C=120$ (panel A), $100$ (B), $80$ (C). The initial load of the units (red dashed line) and the load in the Nash equilibria (red squares) is also reported. The inset displays a magnified plot of the loads, showing that efficient Nash equilibria also induce a decrease in the load on the service units. In panel B  we also report (on a different scale) the frequency of times we found good Nash equilibria in our numerical simulations.}
  \end{center}
\end{figure}

This idea was tested by introducing a modified best response dynamics in which the initial conditions could span the whole spectrum of utilities, generalizing the BR and BRB dynamics studied in the Section \ref{SubSec:dynamics}. The initialization process works as follows: in random order, each user $u$ selects an available service unit $a$ with probability $p_{ua} \propto u_{ua}^{\gamma}$, with $\gamma \in (-\infty,+\infty)$. When all users have attempted to connect to the service system, the configuration is used as the initial condition for the best response dynamics. The standard BR algorithm is obtained for $\gamma=0$, while the BRB one corresponds to the limit $\gamma \to -\infty$. For $\gamma \in (-\infty,+\infty)$,  we can generate configurations with intermediate utility values between the worst and the best ones. Fig.\ref{fig:gammaBR} shows some properties of the equilibria found performing best response from such configurations on instances with the same parameters as in the previous section and no correlation between weights and utilities ($c=0$). For large capacities ($C=120$ in Fig.\ref{fig:gammaBR}a), the best-response dynamics finds Nash equilibria with very high utility (black circles) independently of the utility of the configuration found during the initialization process (black full line). When decreasing the capacity, the best response dynamics start getting trapped in local maxima (Nash equilibria) close to the initial conditions.  Fig.\ref{fig:gammaBR}b shows the case $C=100$, in which a coexistence of ``bad'' and ``good'' equilibria is visible for negative values of $\gamma$, that extends up to $\gamma$ smaller than $4$. The fraction of times the system finds ``bad" Nash equilibria during the dynamical process increases for lower values of $\gamma$ (blue dashed curve in Fig.\ref{fig:gammaBR}b).  This phenomenon becomes dominant when the capacity is further decreased, as shown in Fig.\ref{fig:gammaBR}c for $C=80$, in which all realizations of the dynamics get stuck in the lower-utility Nash equilibria. 
The insets display the average aggregate load on the service units after the initialization process (red full line) and in the Nash equilibria (red squares). It is remarkable that, even though $c=0$,  the more efficient equilibria produce also a slightly lower aggregate load compared to ``bad" equilibria.

We have shown that, at least in the low capacity regime, Nash equilibria with any value of the total utility can be obtained using a simple generalization of the best-response dynamics. In a more general case, ``bad" equilibria still exist, but the system is not sufficiently constrained and best response manages to find the way to the efficient ones. Interestingly, we checked that instead a BP-guided decimation process (this is a process in which iteratively the action of users are fixed following their computed marginals, conditioned to past choices) can be used to find Nash equilibria at almost any value of the utility where they exist, even for large capacity values.

\subsection{Results on stochastic instances}\label{sec:stochastic}
\subsubsection*{Definition of the ensemble} 

The random ensemble of instances is defined as in the deterministic case, with the addition of the probabilities $\{p_u, \, u \in \mathsf{U}\}$ with which user $u$ is active (i.e. participates to the game), which are extracted uniformly at random in the interval $]0,1[$. In the stochastic case, the lower and upper bounds on the total capacity defined in \eqref{eq:C_bounds} must be modified as 
\begin{align} \label{eq:C_bounds_stoch}
  \hat{C}^- &= \sum_u p_u \min_{a \in \dd u} w_{ua} \,, & \hat{C}^+ &= \sum_u p_u \max_{a \in \dd u} w_{ua} \,.
\end{align}
Apart from this (minor) modification, the instance parameters affect the phenomenology of the problem in the same way as in the deterministic case discussed in Section \ref{SubSec:ensemble_det}.

\subsubsection*{Validation of the mirror approximation}

We validated the mirror approximation described in Section \ref{SubSec:mirror} by comparing the average (over the realization of the $t$'s) of the marginals of the variables $y_{ua}$ computed with the mirror approach with the same average marginals computed by sampling explicitly over the realization of the $t$'s. Specifically, we extracted one instance for each value of the parameters we tested, and we converged BP with the mirror fields to compute, for each edge $(ua) \in \mathsf{E}$, the marginal probability $m_{ua} = \mathbbm P[y_{ua} = 2]$ that user $u$ is served by service unit $a$, which is the relevant marginal for the computation of the observables we are interested in. We then extracted, for the same instances, $1\,000$ realizations of the $t$'s, and for each realization computed the same marginal $m_{ua}^{(t)} = \mathbbm P[y_{ua} = 2 | t]$ with an ordinary BP (i.e. without mirror fields), and averaged them over the realizations of the $t$'s, obtaining the average value of the marginal $\overline m_{ua}$ and its standard deviation $\sigma_{ua}$, representing the error on the estimate of $\overline p_{ua}$ due to the finite size of the sampling.

\begin{figure}[h]
\begin{center}    
\includegraphics[width=0.9\columnwidth]{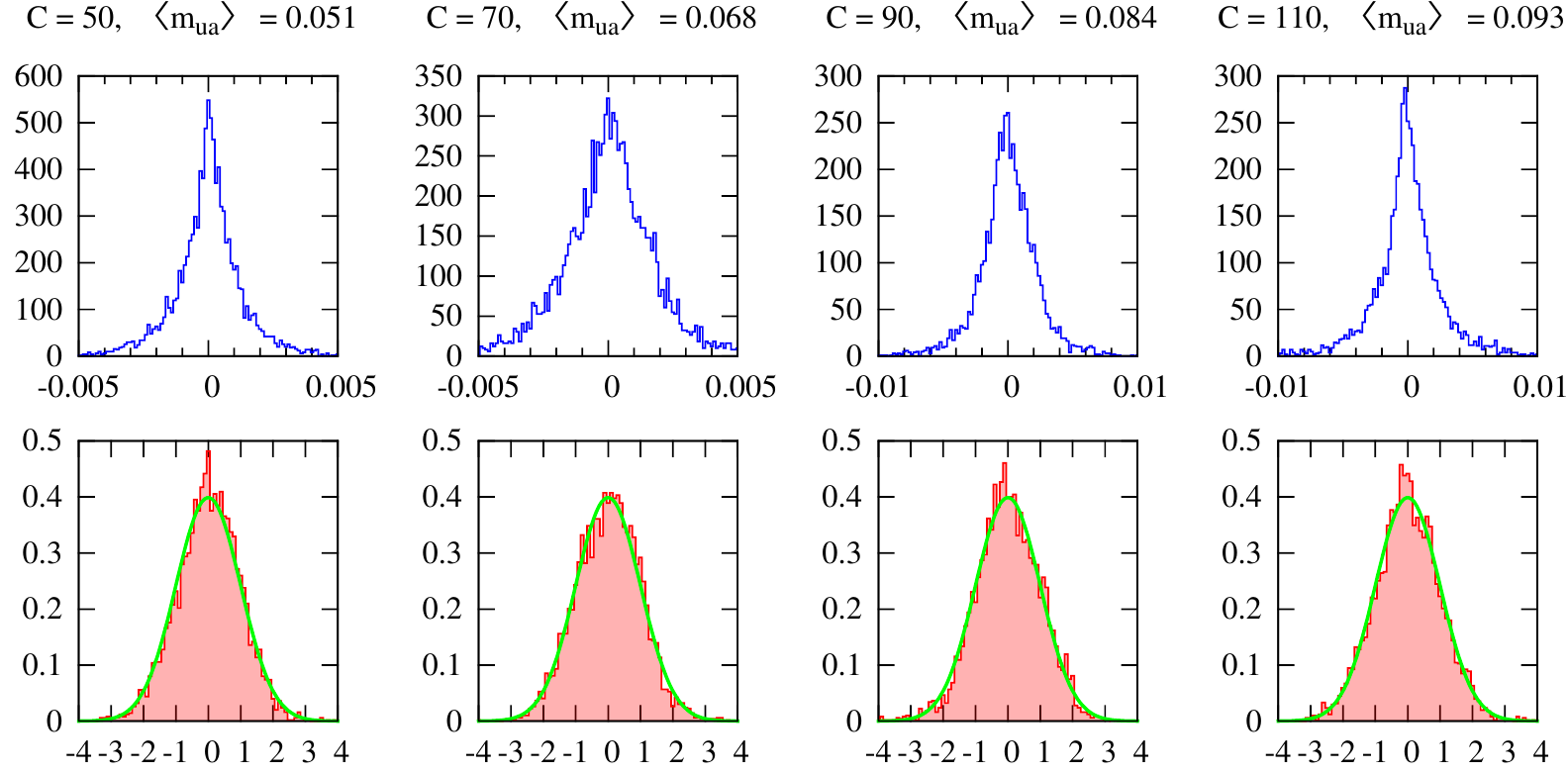}
  \caption{\label{fig:mirror_validation} Distribution of the absolute difference $\Delta_{ua}$ between the mirror and sampling estimates of the marginals (top) and of their normalized error $\delta_{ua} = \Delta_{ua} / \sigma_{ua}$ (bottom) for four individual instances with different capacities $C_a = C \in \{50, 70, 90, 110\}$. The other parameters of the instances are $N = 1\,000$, $M = 50$, $q = 0.1$, $w_\mathrm{min} = 6$, $w_\mathrm{max} = 15$, $v_\mathrm{min} = 1$ and $v_\mathrm{max} = 10$ and $c = 0$, and each $p_u$ is extracted uniformly and independently in $]0,1[$ (these are the values of the parameters we shall consider also in the following Sections). The typical values of the absolute difference are $\sim 2\%$ of the typical values of the marginals $\left< m_{ua} \right>$ for all values of $C$. The distribution of the normalized errors is in excellent agreement with a normal distribution with average 0 and standard deviation 1 (green line in bottom plots), indicating that most of the differences can be attributed to the sampling error.}
  \end{center}
\end{figure}

In Figure \ref{fig:mirror_validation} we show the distributions (over the edges $(ua) \in \mathsf{E}$) of both the absolute difference $\Delta_{ua} = m_{ua} - \overline m_{ua}$ between the mirror and sampling estimates of the marginals, and of their normalized error $\delta_{ua} = \Delta_{ua} / \sigma_{ua}$, for four instances with four different values of the capacity $C$ (which is the most relevant parameter). In all cases, we find that the absolute difference is small compared to the typical value of the marginal $\left< m_{ua} \right> = \sum_{(ua) \in \mathsf{E}} m_{ua} / |\mathsf{E}|$, and that the differences are mainly due to the sampling error.

\subsubsection*{Average values of $U$, $D$ and $C$}

\begin{figure}[h]
\begin{center}
 \includegraphics[width=0.9\columnwidth]{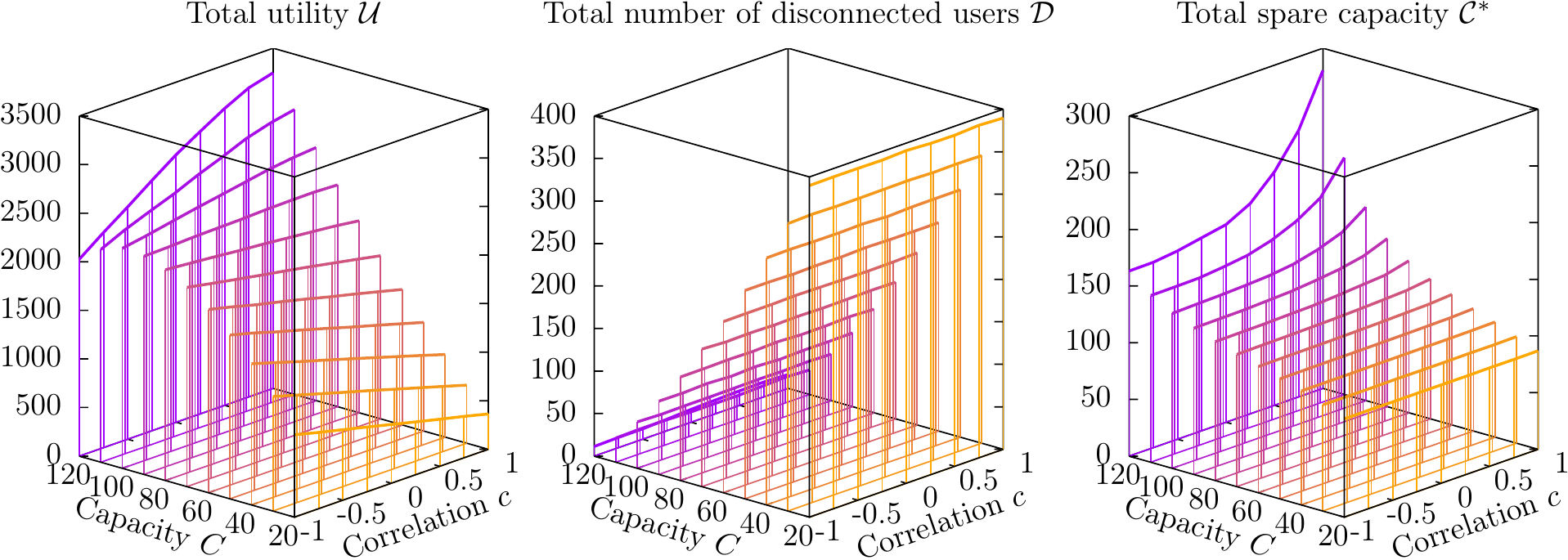}
  \caption{\label{fig:obs_vs_C_and_c_t0} Average values $\mathcal U$, $\mathcal D$ and $\mathcal C^*$ of the observables as a function of the capacity $C_a$ of service units and the correlation $c$ between weight and utility on individual edges. The other parameters are $N=1\,000$, $M=50$, $q=0.1$, $w_\mathrm{min}=6$, $w_\mathrm{max}=15$, $v_\mathrm{min}=1$ and $v_\mathrm{max}=10$. Each data point, corresponding to a vertical line, is an average over $240$ instances for $C \leq 100$, and over $40$ to $120$ instances for $C \geq 110$, and the standard deviations are of the order of the width of the lines. For $C = 120$ a significant number of instances did not converge, and where therefore excluded from the dataset.}
  \end{center}
\end{figure}

As for the deterministic case, we study the average values of the relevant observables as a function of the capacity $C$ of individual service units and of the correlation $c$ between the weight $w_{ua}$ and the utility $u_{ua}$ on each edge $(ua) \in \mathsf{E}$. Once the range of capacities is rescaled to take into account both the smaller number of service units ($M = 50$ vs. $M = 200$ in the deterministic case) and the fact that the expected number of \emph{active} users is $\overline N = 500$ (whereas all $N = 1\,000$ users are active in the deterministic case), we recover exactly the same phenomenology we observed in the deterministic case, and we refer the reader to Section \ref{SubSec:dynamics} for a discussion of the qualitative features of the numerical results shown in Figure \ref{fig:obs_vs_C_and_c_t0}.

\subsubsection*{Comparison with explicit dynamics}

\begin{figure}[h]
\begin{center}
  \includegraphics[width=0.9\columnwidth]{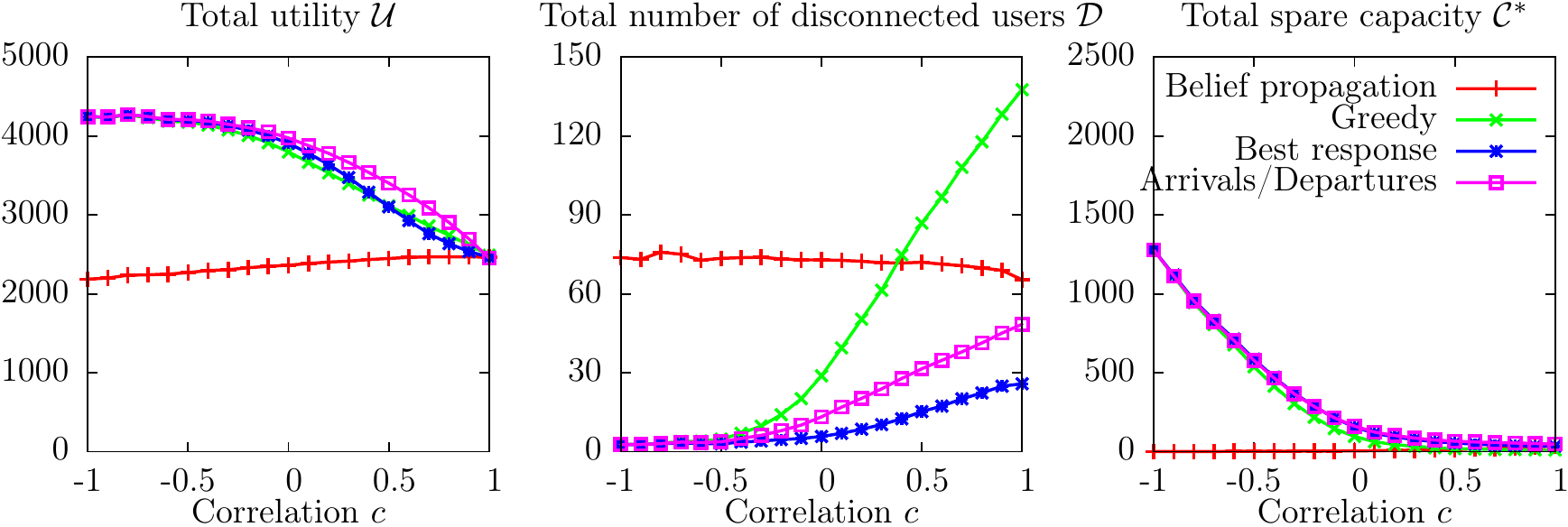}
  \caption{\label{fig:t0_UDC_vs_c} Average values $\mathcal U$, $\mathcal D$ and $\mathcal C^*$ of the observables as a function of the correlation $c$ between weight and utility on individual edges. The other parameters are $N=1\,000$, $M=50$, $C=50$, $q=0.1$, $w_\mathrm{min} = 6$, $w_\mathrm{max} = 15$, $v_\mathrm{min} = 1$ and $v_\mathrm{max} = 10$. Each data point is an average over $\sim 70$ instances, and for each instance, each dynamics is realized $30\,000$ times. The average value of each observable is computed over the instances and over the realizations. The standard deviations (of the averages over the realizations of the dynamics across different instances) are much smaller than symbol sizes, while the standard deviations (of the averages computed with Belief Propagation over the instances) is of the order of the symbol sizes.}
  \end{center}
\end{figure}

As for the deterministic case, we compare the average values of the observables obtained by averaging uniformly over all Nash equilibria with Belief Propagation with the results of the explicit simulation of three dynamics which converge to Nash equilibria:
\begin{itemize}
 \item \emph{Greedy (G)} -- A list of active users is extracted based on the probabilities $\{p_u\}$ that each user $u$ is active, and then we proceed as in the deterministic case for the active users
 \item \emph{Best Response (BR)} -- As for Greedy, we first extract a list of active users and then proceed as in the deterministic case for the active users
 \item \emph{Arrivals/Departures (A/D)} -- We consider a discrete time dynamics in which at each time step we extract a random permutation of all users, and then in the order of the permutation each active user becomes inactive with probability $(1-p_u)/N$ while each inactive user becomes active with probability $p_u/N$ and selects the best available service unit in a greedy way. At the end of each time step, Best Response is run until convergence for all active users in order to reach an equilibrium. The dynamics is repeated for a fixed number of time steps ($100 \, N$ in the numerical simulations).
\end{itemize}
Numerical results for the comparison are shown in Figure \ref{fig:t0_UDC_vs_c}. As in the deterministic case, BP gives results that are very different from the three dynamics, which appear to be strongly biased towards ``good'' equilibria. Whereas in the deterministic case the BRB dynamics gives results that are significantly different from the other two (at least for some values of $c$), the A/D dynamics gives results which are quite similar to both G and BR, except for the number of disconnected users $\mathcal D$ when $c \geq -0.5$. A detailed analysis on single instance following the steps of Section \ref{SubSec:entropy} was not performed for the stochastic case.

\section*{Discussion}

We proposed a simple game theoretic model of distributed service provision, in which users want to be served by the service unit they prefer and they indirectly interact because of a capacity constraint. The existence of at least one Nash equilibrium is guaranteed by the fact that the game belongs to the class of potential games. The potential function is the total utility of the users and, for construction,  the best response dynamics always converge to a local maximum of the potential, which is also a Nash equilibrium. The Nash equilibrium conditions can be rephrased as a set of hard constraints on configurations of binary variables (the choices of the users) and analyzed using standard statistical mechanics methods, such as belief propagation and message-passing algorithms. Moreover, a soft constraint in the form of an energetic term can be included in the factor graph formulation in order to study the properties of Nash equilibria with given average total utility. We derived belief propagation equations for this problem and studied the properties of the corresponding Nash equilibria on random instances of the service provision game, in which the user connectivity to the service units follows a Poisson distribution while the load weight provided by the users to the service units and the corresponding payoffs are drawn from a uniform distribution. The analysis of the Nash equilibrium landscape reveals a  large variety of equilibria, with very different total utility. Best-response dynamics from random initial conditions usually tend to large-utility equilibria, even though those of smaller utility are exponentially more numerous. In order to prove that these equilibria can be actually reached by a simple dynamical process, we modified the best-response dynamics initializing it to a set of non-random initial configurations. These configurations are selected to be close to the saturation limit of the service unit capacities. In this regime, the rearrangement induced by the best response is very limited and the Nash equilibria obtained have a total utility close to the one of the initial configurations. 

Other interesting phenomena appears when utility values and weights are correlated. The average total utility increases when the correlation $c$ between weights and utility values increases, that is when the competition between users becomes harsher, whereas the opposite is observed in the low capacity regimes. 
Moreover, quite surprisingly the average spare capacity of Nash equilibria seems to increase with the correlation in the large capacity region. This means that, even if the users tend to choose the units on which they bring larger loads, they do it in a very optimized way.     
Our characterization of the equilibria also makes possible to estimate the price of anarchy of the game, which decreases smoothly from increasing the correlation. 

In many realistic situations, the instance of the game theoretic problem changes over time, because the agents could follow very complex temporal activity patterns. In the absence of any precise information on the dynamics of the agents, the complexity of agents' activity is summarized into a set of stochastic parameters $\{t_u\}$, that indicate whether user $u$ participate or not to the game. Users play a game on the deterministic instance corresponding to a single realization of the stochastic parameters, then the equilibrium properties should be averaged over different realizations of the parameters. Our results show that the average properties of the Nash equilibria in the stochastic case are qualitatively similar to those observed in the case of deterministic instances. However, our results are very relevant from a methodological viewpoint. Instead of resorting to sampling techniques, we perform the average over fixed realizations of the stochastic parameters (quenched average) by means of a systematic approximation scheme that, at least at the first-order level, can be naturally incorporated in the belief propagation approach, at the cost of  introducing some additional messages, that we call ``mirror messages". In the case under study, the method provides a very accurate estimate of all the average properties of interest. We believe that this approach could be useful in several problems in which it is necessary to perform averages over fixed realizations of disordered parameters, whenever the correlations between variables in the system are not too strong. 

\section*{Acknowledgments}We are grateful to R. Zecchina for useful discussions. We acknowledge the European Grant ERC No. 267915 and Italian FIRB Project No. RBFR10QUW4. 


\end{document}